\documentclass[11pt]{article}
\usepackage[utf8]{inputenc}
\usepackage{amsmath}
\usepackage{graphicx}
\usepackage{enumerate}
\usepackage{natbib}
\usepackage[english]{babel}
\usepackage{amsfonts}
\usepackage{amsthm}
\graphicspath{ {Figures/} }
\usepackage{setspace}
\usepackage[table, dvipsnames]{xcolor}
\usepackage{algorithm}
\usepackage{algpseudocode}

\newtheorem{theorem}{Theorem}

\newcommand{\betavec}{\boldsymbol{\beta}}

\newcommand{\zerovec}{\boldsymbol{0}}
\newcommand{\onevec}{\boldsymbol{1}}

\newcommand{\dvec}{\boldsymbol{d}}
\newcommand{\evec}{\boldsymbol{e}}

\newcommand{\hvec}{\boldsymbol{h}}

\newcommand{\yvec}{\boldsymbol{y}}

\newcommand{\A}{\textbf{A}}
\newcommand{\C}{\textbf{C}}
\newcommand{\D}{\textbf{D}}

\newcommand{\Hmat}{\textbf{H}}
\newcommand{\I}{\textbf{I}}

\newcommand{\Kmat}{\textbf{K}}
\newcommand{\M}{\textbf{M}}

\newcommand{\Pmat}{\textbf{P}}

\newcommand{\V}{\textbf{V}}

\newcommand{\X}{\textbf{X}}

\newcommand{\blind}{1}

\begin{document}

\def\spacingset#1{\renewcommand{\baselinestretch}%
{#1}\small\normalsize} \spacingset{1}


\if1\blind
{\title{\bf Powerful Foldover Designs}
  \author{Jonathan Stallrich\\ 
  NCSU, Statistics, jwstalli@ncsu.edu\\
  Rakhi Singh\\
    Indian Institute of Technology Madras, Mathematics, \\rakhi@smail.iitm.ac.in\\
Kyle Vogt-Lowell\\ 
NCSU, Chemical and Biomolecular Engineering,\\ kjvogtlo@ncsu.edu\\
 Fanxing Li\\
 NCSU,
 Chemical and Biomolecular Engineering\\
 fli5@ncsu.edu}
  \maketitle
} \fi

\if0\blind
{
  \bigskip
  \bigskip
  \bigskip
  \begin{center}
    {\LARGE\bf Foldover Designs for Inference and Prediction of second order Models}
\end{center}
  \medskip
} \fi

\bigskip
\begin{abstract}
\noindent 
The foldover technique for screening designs is well known to guarantee zero aliasing of the main effect estimators with respect to two factor interactions and quadratic effects. It is a key feature of many popular response surface designs, including central composite designs, definitive screening designs, and most orthogonal, minimally-aliased response surface designs. In this paper, we show the foldover technique is even more powerful, because it produces degrees of freedom for a variance estimator that is independent of model selection. These degrees of freedom are characterized as either pure error or fake factor degrees of freedom. A fast design construction algorithm is presented that minimizes the expected confidence interval criterion to maximize the power of screening main effects. An augmented design and analysis method is also presented to avoid having too many degrees of freedom for estimating variance and to improve model selection performance for second order models. Simulation studies show our new designs are at least as good as traditional designs when effect sparsity and hierarchy hold, but do significantly better when these effect principles do not hold. A real data example is given for a 20-run experiment where optimization of ethylene concentration is performed by manipulating eight process parameters.


\end{abstract}

\noindent
{\it Keywords: Expected confidence interval criterion, half design, pre-selection variance estimator, pure error, response surface methodology, screening experiment
}\\

\noindent
We confirm that there are no relevant financial or non-financial competing interests to report.\\

\noindent
Funding: No funding was received.

\vfill

\newpage
\spacingset{1.45} 





\maketitle

\section{Introduction}

Response surface methodology sequentially optimizes a process by manipulating $m$ factors. The first step is a screening experiment to determine which $k < m$ factors drive most of the response variation. By performing a screening experiment in a localized region, one may assume a second order model comprised of an intercept, linear main effects, bilinear interactions, and quadratic main effects. There are $1+2m+\binom{m}{2}$ such effects and a large experiment is needed to estimate all of them. The effect principles described in \cite{WHtextbook} usually govern screening experiments where the goal is to determine which $k \leq m$ linear main effects are most significant. This objective justifies performing a small screening experiment focusing on optimizing statistical inference properties for the main-effects-only model. However, ignoring second order terms can bias main effect inferences. This bias can be eliminated by constructing designs with the foldover technique, which flips the signs of the settings of an initial design and appends these new runs to the initial design. The foldover technique has been primarily employed for two-level designs \citep{Diamond1995,Miller01022001,Li01052003,Miller01112005,LIN20083107,Errore02012017,Nguyen02012020}, but has recently found new uses with definitive screening designs \citep{jones2011class} and most orthogonal, minimally-aliased response surface designs \citep{OMARS}
having numeric factors. The goal of this paper is to highlight and demonstrate that the foldover technique is even more powerful because it can also produce an unbiased estimator for error variance, $\sigma^2$, which is vital for reliable main effect inferences and process optimization.

This paper was motivated by a collaborative effort by the authors to 
determine the relationship between $m=8$ process parameters (see Table~\ref{tab:EthaneSettings}) and seven reactive performance outcomes of a novel molten-salt-mediated scheme for sustainable ethylene 
production from ethane
. Understanding these relationships would serve to inform operating conditions employed in a commercial-scale process model and indirectly probe underlying reaction mechanisms. 
Resource constraints permitted only $n=20$ total runs. Due to \textit{a priori }uncertainty surrounding the number and magnitude of interactions in the model, the methodology in this paper was used to generate a design with zero aliasing that produced an unbiased estimator of $\sigma^2$ to enhance the reliability of the results.
\begin{table}
    \centering
     \caption{Ethylene experiment's factors and their ranges.}
    \begin{tabular}{|l|c|c|} \hline 
         Factor (units) & Low Level & High Level\\ \hline 
          Total gas flow rate (SCCM) & 25 &50 \\ \hline 
          Temperature (°C) &  750 & 800\\ \hline 
          Reduction step duration (minutes) &  5  &15 \\ \hline 
          Ethane feed concentration (\%) & 15 &75 \\ \hline 
          Regeneration step duration (minutes) &  2  &10 \\ \hline 
          Carbonate sample loading (g)  &  10 &30 \\ \hline 
          CO\textsubscript{2} feed concentration (\%)  &  3 &15 \\ \hline 
          Gas injection orifice size (inches)  &  2/16 &3/16 \\ \hline
    \end{tabular}
    \label{tab:EthaneSettings}
\end{table}

\subsection{Background}\label{sec:Background}
The $i$-th run of an experiment declares settings for the $m$ factors of interest, represented by $\dvec_i^T=(d_{i1},\dots,d_{im})$, and yields a response $y_i$. Each factor is either numeric with $d_{ij} \in [-1,1]$ (after proper scaling) or categorical with $d_{ij}=\pm 1$. The second order model has the form
\begin{align}
    y_i = \beta_0 + \sum_{j=1}^m d_{ij}\beta_j + \sum_{j<j'} d_{ij}d_{ij'}\beta_{jj'} + \sum_{j=1}^m d_{ij}^2\beta_{jj}+e_i\ ,\ \label{eqn:QuadMod_Scalar}
\end{align}
where $e_i\sim^{iid}N(0,\sigma^2)$. The linear main effects, bilinear interaction effects, and quadratic effects correspond to the $\beta_j$, $\beta_{jj'}$, and $\beta_{jj}$, respectively. The two-factor interaction model assumes all $\beta_{jj}=0$ while the quadratic model includes all terms in \eqref{eqn:QuadMod_Scalar}. Note that categorical factors do not have a quadratic effect. The matrix form of \eqref{eqn:QuadMod_Scalar} for an $n \times m$ design matrix $\D$ with rows $\dvec_i^T$ is
\begin{align}
    \yvec=\X\betavec+\evec=\X_1 \betavec_1 +\X_2 \betavec_2 + \evec\ ,\
    \label{eqn:QuadMod_Matrix}
\end{align}
where $\X_1=(\onevec \, | \, \D)$ is the main effect model matrix and $\X_2$ is the model matrix for the second order terms, having columns with elements $d_{ij}d_{ij'}$ for the two-factor interaction model and  $m$ additional columns with elements $d_{ij}^2$ for the quadratic model. 

The least-squares estimator $\hat{\betavec}=(\X^T\X)^{-1}\X^T\yvec$ is unbiased when the information matrix, $\X^T\X$, is full rank. For designs with small $n$, $\X^T\X$ is not invertible, so one may instead try to estimate the main effects using  $\hat{\betavec}_1=(\X_1^T\X_1)^{-1}\X_1^T\yvec$. We want to identify a design whose $\hat{\betavec}_1$ is close to $\betavec_1$ with high probability. For a fixed $\D$, $\hat{\betavec}_1$ is normally distributed with variance $\text{Var}(\hat{\betavec}_1)=\sigma^2(\X_1^T\X_1)^{-1}$. The diagonal entries of $(\X_1^T\X_1)^{-1}$, denoted by $v_0,v_1,\dots,v_m$, are called the \emph{design variances} under the main effect model. Designs that minimize the design variances are preferred, but we must also concern ourselves with the potential bias, $\text{E}(\hat{\betavec}_1-\betavec_1)=(\X_1^T\X_1)^{-1}\X_1^T\X_2 \betavec_2$ under model~\eqref{eqn:QuadMod_Matrix}. The bias depends on the unknown $\betavec_2$ unless $\A \equiv (\X_1^T\X_1)^{-1}\X_1^T\X_2=\zerovec$, where $\A$ is called the alias matrix. Designs that simultaneously minimize the design variances and biases of $\hat{\betavec}_1$ have been well studied \citep{GMA,G2, cox2000theory,Min_G_ab,WHtextbook,Jones_2011,mead2012statistical}. The traditional approach involves finding an orthogonal main effect plan, having the minimum $\text{Var}(\hat{\betavec}_1) = \sigma^2/n \ \I$, that also minimizes $\A$ in some sense (e.g.,  $\text{tr}(\A^T\A)$, minimum aberration, generalized aberration). 

For reliable main effect inferences, it is also critical to have an unbiased estimator of $\sigma^2$.  
The default estimator $\hat{\sigma}^2_{X_1}=\yvec^T(\I-\Pmat_{X_1})\yvec/(n-m-1)$, where $\Pmat_{X_1}=\X_1(\X_1^T\X_1)^-\X_1^T$ is the orthogonal projection onto the column space of $\X_1$ with $\M^-$ denoting a generalized inverse of $\M$, is biased when $\betavec_2 \neq \zerovec$, causing $\text{E}(\hat{\sigma}^2_{X_1}) > \sigma^2$. Some analysis methods yield an unbiased estimator after performing model selection, but an incorrect selection can again lead to a biased estimator. \cite{Stallrich07082024} recommend the \emph{pre-selection estimator}, $\hat{\sigma}^2_{X}=\yvec^T(\I-\Pmat_{X})\yvec/(n-\text{rank}(\X))$, which is unbiased when $\text{rank}(\X)<n$.



The $g_X=n-\text{rank}(\X)$ degrees of freedom for $\hat{\sigma}^2_{X}$ can be partitioned into two components. The first component is the $p$ pure error degrees of freedom due to replicated rows in $\D$. If $\dvec_i$ appears $n_i$ times in $\D$, there will be $n_i-1$ pure error degrees of freedom. The name comes from the fact that the pure error variance estimator, $SS_{pe}/p$ where $SS_{pe}$ is the pure error sum of squares, is always unbiased regardless of the true model. The remaining $\ell_X= (g_X-p)$ degrees of freedom are called lack of fit degrees of freedom which produce the lack of fit variance estimator, $SS_{LOF}/\ell_X$. Unlike $SS_{pe}/p$, $SS_{LOF}/\ell_X$ is a biased estimator of $\sigma^2$ if the fitted model is underspecified. This bias could, for example, result from the presence of other interaction effects (e.g. $d_1d_2^2$ or $d_1d_2d_3$) that are ignored in model~\ref{eqn:QuadMod_Scalar}. Unless otherwise specified, we will assume $\text{E}(SS_{LOF}/\ell_X)=\sigma^2$, giving the expression
\begin{align}
    \hat{\sigma}^2_X=\frac{SS_{pe}+SS_{LOF}}{p+\ell_X}\ .\
\end{align}

Recent papers have emphasized constructing designs that have a well-defined $\hat{\sigma}^2_X$. \cite{gilmour2012optimum} combined minimizing the $D$- and $A$-criterion while requiring a given $g_X$, but they ignored the need to minimize potential bias. \cite{leonard2017bayesian} remedied this by replacing the $D$-criterion with its Bayesian version, following \cite{dumouchel_jones1994}. \cite{jones2017effective} augmented definitive screening designs (DSDs) by adding $f$ fake factors to the generating conference matrix, thereby adding $2f$ runs to the original DSD construction and producing $\ell_X=f$ lack of fit degrees of freedom. Partial replication was also introduced to these augmented DSDs in \cite{jones2020partial}. 

Assuming a distribution on the second order effects, $\betavec_2\sim N(0,\tau^2 \I)$, \cite{Stallrich07082024} proposed a more flexible optimality criterion based on minimizing the average maximum deviation of so-called expected confidence intervals (ECI) from the true signal to noise ratio $\beta_j/\sigma$:
\begin{align}
\mbox{ECI} = \frac{1}{m} \sum_j \left\{\sqrt{\frac{2\tau^2}{\pi}\A_j^T\A_j} + c(g_X)t_{\alpha/2,g_X} \sqrt{v_j}\right\}\ ,\ \label{eqn:ECI}
\end{align}
where $\A_j^T$ is the $j$-th row of $\A$; $c(g_X) > 1$ is a constant with $c(g_X) \to 1$ as $g_X \to \infty$; and $t_{\alpha/2,g_X}$ is the $\alpha/2$ upper tail critical value for a $t$-distribution with $g_X$ degrees of freedom. Each term in the sum represents the maximum deviation of the $j$-the confidence interval from the true $\beta_j/\sigma$. If the maximum deviation is larger than $\beta_j/\sigma$, we would expect the confidences intervals to contain 0. If $\beta_j > 0$, we would then likely fail to identify the factor as active. Therefore, the ECI criterion can be interpreted as the smallest average $\beta_j/\sigma$ that can be reliably detected in the first stage. The design that minimizes the ECI criterion will be more powerful in detecting active effects. The main benefit of the ECI criterion is its ability to balance three different objectives (minimizing design variances, minimizing aliasing, and maximizing $g_X$) according to their relative contributions to the desired inference. The main drawback, however, is that their design construction algorithm can be inefficient, and does not easily allow the user to specify a desired $\ell_X$ and $p$. These drawbacks, however, are nearly eliminated in this article by focusing on foldover designs.

\subsection{Foldover Designs}
A foldover design, $\D^\pm$, can be represented by
\begin{align}
\D^{\pm}=\begin{pmatrix}\phantom{-}\Hmat \\ -\Hmat\end{pmatrix} \ ,\  \label{eqn:foldover} 
\end{align} 
where $\Hmat$ is a \emph{half design} matrix with $n/2$ runs and $\text{rank}(\Hmat)=m$. Throughout this article, designs that are not a foldover will be denoted by $\D$, but it is possible for a subset of the rows of $\D$ to be a foldover design.  The identities $d_{ij}d_{ij'}=(-d_{ij})(-d_{ij'})$ and $(-d_{ij})^2=d_{ij}^2$ make the second order model matrix for $\Hmat$, denoted $\X_{H2}$, equal to the second order model matrix for $-\Hmat$. The information matrix for $\betavec$ under $\D^\pm$ then admits the expression
\begin{align}
    2\begin{pmatrix}
    n/2 & \zerovec &  \onevec^T\X_{H2}\\
    \zerovec &  \Hmat^T\Hmat & \zerovec\\
     \X_{H2}^T\onevec & \zerovec &  \X_{H2}^T\X_{H2} 
\end{pmatrix}\ ,\ \label{eqn:Foldover_InfoMat}
\end{align}
making the $m$ main effect estimators in $\hat{\betavec}_1$ unbiased with $\text{Var}(\hat{\betavec}_1)=\sigma^2/2(\Hmat^T\Hmat)^{-1}$.  

The foldover technique is a key ingredient for many response surface designs. Resolution IV fractional factorial designs can be constructed by folding over a resolution III design \citep[][page 310]{Montgomery2019}. Central composite designs \citep[][page 248]{Montgomery2019}, or CCDs, are comprised of three components: (a) a two-level (fractional) factorial design, (b) $n_0$ center runs, and (c) $2m$ axial runs comprised of the $m$ design points $\alpha\I_m$ and their foldover $-\alpha\I_m$. A foldover design could also be chosen for (a) to give desirable analysis properties of the main effects. The original construction of DSDs in \cite{jones2011class} also involved an algorithmic search of a $D$-optimal design of an $m \times m$ half design with the diagonal elements fixed to 0. The design was then folded over and a center run was appended. \cite{xiao2012} noted that for even $m$, one could employ an order-$m$ conference matrix, $\C_m$, for the half design which has the same structure and has $\C_m^T\C_m=(m-1)\I_m$, giving orthogonal main effect estimators. The orthogonality structure from this construction was embraced by \cite{OMARS} to create OMARS design with more flexible run sizes; most OMARS designs are foldover designs.

The recommended analysis of foldover designs generally follows a two-stage approach, with the first stage estimating the set of active factors under the main effect model \citep{Miller01112005}. Following the effect heredity principle, model selection inference is then performed on the second order effects involving the estimated active factors. For example, \cite{jones2017effective} recommend a guided subsets approach, and \cite{Stallrich07082024} propose an all-subsets approach with model selection via a modified BIC criterion. However, poor screening in the first stage will hinder the capabilities of the second stage analysis, so it is reasonable to focus on identifying designs that optimize the inference properties of the first stage. 

\subsection{Contributions and Paper Overview}
This paper reverses the traditional approach of optimal screening designs by finding a foldover design (which guarantee $\A=\zerovec$) that minimizes $\text{Var}(\hat{\betavec}_1)$ and has a well-defined $\hat{\sigma}^2_X$. To this end, Section~\ref{sec-2} provides a complete characterization of the $g_X$ degrees of freedom for a foldover design, which, to the best of our knowledge, has not appeared in the literature. Remaining contributions utilize this characterization to construct powerful foldover designs for different scenarios. Section~\ref{sec:alg} starts with a general algorithmic construction of foldover designs that modifies the ECI algorithm by \cite{Stallrich07082024}. It then describes specialized construction techniques applied to the case of two-level designs.  Section~\ref{sec:FoldoverAug} introduces a design augmentation approach that flexibly adds runs to a foldover design under a Bayesian $A$ criterion focused on estimating second order effects. This approach is needed when $n$ is odd and is recommended when $n$ is relatively large, lest the foldover design becomes overpowered for the main effects and loses its ability to identify important second order effects. The augmented structure yields a straightforward partitioned analysis method when the augmented $\D$ has $\A\neq \zerovec$. Section~\ref{sec:Sim} provides two examples demonstrating the construction methods and the analysis methods via a simulation study. Section~\ref{sec-expt} presents the results of the recommended foldover design implemented for the Ethylene motivating example, and we conclude the article with a discussion in Section~\ref{sec-disc}. 

\section{Estimating $\sigma^2$ with Foldover Designs{\label{sec-2}}}
The primary contribution of this article is the characterization of a foldover design's $g_X$ degrees of freedom for $\hat{\sigma}^2_X$ under a second order model. To provide insight into the characterization, consider the following three half designs with $m=4$ factors and $8$ runs, where levels $1$ and $-1$ are represented by $+$ and $-$, respectively:
\begin{align}
\Hmat_1=\left(\begin{array}{cccc}
+ & + & + & +\\ \hline
- & + & - & +\\ \hline
+ & - & - & +\\ \hline
- & - & + & +\\ \hline
+ & + & + & -\\ \hline
- & + & - & -\\ \hline
+ & - & - & -\\ \hline
- & - & + & -
\end{array}\right) \quad 
\Hmat_2=\left(\begin{array}{cccc}
0 & 0 & 0 & 0\\ \hline
- & + & - & +\\ \hline
+ & - & - & +\\ \hline
- & - & + & +\\ \hline
+ & + & + & -\\ \hline
- & + & - & -\\ \hline
+ & - & - & -\\ \hline
- & - & + & -
\end{array}\right) \quad 
\Hmat_3=\left(\begin{array}{cccc}
+ & - & - & +\\ \hline
+ & + & + & +\\
+ & + & + & +\\
\hline
+ & - & + & -\\
+ & - & + & -\\
- & + & - & +\\
\hline
+ & + & - & -\\
- & - & + & +
\end{array}\right) \ .\
\label{eqn:Section2_HalfDesigns}
\end{align}
The runs are partitioned into groups using the horizontal lines. The runs in a group correspond to replicating or folding over a unique row, $\hvec_g$. For example, all eight groups for $\Hmat_1$ and $\Hmat_2$ include one replicate of a unique row. Note that $\Hmat_2$ is nearly identical to $\Hmat_1$, but with the first row equal to a center run. The half design $\Hmat_3$ has four groups where group $1$ has one replicate of $\hvec_1^T=(+, -, -, +)$; group $2$ has $2$ replicates of $\hvec_2^T=(+, +, +, +)$; group $3$ has two replicates of $\hvec_3^T=(+, -, +, -)$ and one foldover row; and group $4$ one replicate of $\hvec_4^T=(+, +, -, -)$ and one foldover row.     

The three corresponding foldover designs, denoted $\D_1^\pm$, $\D_2^\pm$, and $\D_3^\pm$, have $n=16$ runs which can also be partitioned into the same groups but with twice as many runs in each group. Foldover design $\D_1^\pm$ will comprise eight groups, each having a foldover pair. All of its runs are unique so there are $p=0$ pure error degrees of freedom. Similar to $\D_1^\pm$, there are also eight groups of runs for $\D_2^\pm$, but now the first group has two replicates of the center run, and the other seven groups comprise of foldover pairs. Hence, $\D_2^\pm$ has $p=1$ pure error degree of freedom. The four groups in $\D_3^\pm$ are more interesting. Group 1 has two runs that are a foldover pair of $\hvec_1$ and so does not contribute pure error degrees of freedom. Group 2 replicates both $\hvec_2$ and $-\hvec_2^T=(-,-,-,-)$ twice, producing two pure error degrees of freedom.  Group 3 of $\D_3^\pm$ replicates both $\hvec_3$ and $-\hvec_3^T=(-,+,-,+)$ three times, giving $(3-1)+(3-1)=4$ pure error degrees of freedom. Finally, group 4 of $\D_3^\pm$ replicates $\hvec^T_4$ and $-\hvec^T_4=(-,-,+,+)$ twice, giving 2 pure error degrees of freedom. Therefore, $\D_3^\pm$ has $p=2+4+2=8$ pure error degrees of freedom.

Recall the lack of fit degrees of freedom, $\ell_X$, generally depends on the chosen second order model. However, some of the lack of fit degrees of freedom for a foldover design can be generated by the half design and are independent of the second order model. We will call these \textit{fake factor} degrees of freedom due to their following construction. To demonstrate, $\text{rank}(\Hmat_1)=m$ and has $8$ rows. Therefore, there exists a matrix $\V_1$ with $4$ orthonormal columns where $\Hmat_1^T\V_1=\zerovec$. If we append these $\V_1$ columns to $\Hmat_1$, they could be viewed as settings for so-called fake factors (i.e., factors that are not actually manipulated in the experiment). It follows then that the columns in the foldover of $\V_1$, denoted $\V_1^\pm$, will be orthogonal to every column in $\X$ for \textit{any} second order model matrix. Therefore, the corresponding sum of squares should be included in $\hat{\sigma}^2_X$. Since $\D_1^\pm$ has no replicated runs, it follows that these sum of squares must be part of $SS_{LOF}$. Therefore, we say that $\Hmat_1$ generates a foldover design with $f=4$ fake factor degrees of freedom.

Describing fake factor degrees of freedom takes a bit more work when the half design includes center runs, replicated runs, or foldover runs. To demonstrate, both $\Hmat_2$ and $\Hmat_3$ have rank $4$ with $8$ runs and, following the same reasoning as $\Hmat_1$, producing fake factor matrices $\V_2$ and $\V_3$. However, both half designs have runs that, when folded over, produce pure error degrees of freedom. This causes some of the columns in $\V_2$ and $\V_3$ to correspond to pure error degrees of freedom. Our definition requires fake factor degrees of freedom to be part of the lack of fit degrees of freedom. For example, the center run for $\Hmat_2$ causes $p=1$, and leads to $(1, 0, \dots, 0)^T$ as a column in $\V_2$. Therefore, it should not be considered as a fake factor column, leaving $f=3$ fake factor degrees of freedom. For $\Hmat_3$, each group with $n_g > 1$ runs produces $n_g-1$ columns in $\V_3$ that correspond to pure error degrees of freedom (see the Supplementary Materials). These columns should also not be counted as fake factor columns. Therefore, half design $\Hmat_3$ generates a foldover design with $f=4-(2-1)-(3-1)-(2-1)=4-4=0$ fake factor degrees of freedom.

We now present Theorem~\ref{thm:MainResult} which generalizes the findings from the previous example:
\begin{theorem}
Let $\Hmat$ be a $(m+v) \times m$ half design where $\text{rank}(\Hmat)=m$.  Partition $\Hmat$ into $G+1$ distinct groups of runs with group $g$ represented by a unique row $\hvec_g$ where
\begin{itemize}
    \item Group $g=0$ includes all $n_0 \geq 0$ center runs (i.e., $\hvec_0=\zerovec$). This group may be empty.
    \item Group $g=1,\dots,G$ includes all $n_g \geq 1$ rows equal to $\hvec_g$ or its foldover, $-\hvec_g$. 
\end{itemize}
Then, under any second order model, the foldover design $\D^{\pm}$ has fake factor and pure error degrees of freedom equal to
\begin{align}
    f &=v-n_0-\sum_{g=1}^G (n_g-1)\ ,
    \ \label{eqn:Thm1_f}\\
    p &=\max(0,2n_0-1)+2\sum_{g=1}^G (n_g-1) \label{eqn:Thm1_p}\ .\
\end{align}\label{thm:MainResult}
\end{theorem}
\noindent
A formal proof of Theorem~\ref{thm:MainResult} is provided in the Supplementary Materials. 

Table~\ref{tab:Example1} summarizes the properties of these three designs using the notation in the theorem. Note that $\D^\pm_1$ has $f=4$ fake factor degrees of freedom, as discussed earlier, but also has an additional lack of fit degree of freedom. This means there exists some vector that is orthogonal to the design's model matrix as well as $\V_1^\pm$. A similar phenomenon happens with $\D_2^\pm$, but the additional lack of fit degree of freedom only exists for the two-factor interaction model.

\begin{table}[H]
    \centering
    \caption{Properties of foldover designs from the half designs $\Hmat_1$, $\Hmat_2$, and $\Hmat_3$ from \eqref{eqn:Section2_HalfDesigns}.}
    \begin{tabular}{c|ccccccccc|c c|c c|c c}
       \multicolumn{10}{c}{} & \multicolumn{2}{c}{} & \multicolumn{2}{c}{2FI} & \multicolumn{2}{c}{Quad}\\
       Design  & $n_0$ & $n_1$ & $n_2$ & $n_3$ & $n_4$ & $n_5$ & $n_6$ & $n_7$ & $n_8$ & $f$ & $p$ & $\ell_X$ &  $g_X$ & $\ell_X$ &  $g_X$ \\ \hline
       $\D_1^\pm$ & 0 & 1 & 1 & 1 & 1 & 1 & 1 & 1 & 1 & 4 & 0 & 5 & 5 & 5 & 5 \\
       $\D_2^\pm$ & 1 & 1 & 1 & 1 & 1 & 1 & 1 & 1 &   & 3 & 1 & 4 & 5 & 3 & 4 \\
        $\D_3^\pm$ & 0 & 1 & 2 & 3 & 2 &   &   &   &   & 0 & 4 & 4 & 8 & 4 & 8 \\
    \end{tabular}
    \label{tab:Example1}
\end{table}

One immediate consequence is that the original construction of DSDs will have $g_X=0$ because the half design has $v=0$. This explains why \cite{jones2017effective} needed to augment the half design with more runs. Another consequence is that constructing foldover designs for large $n$ may not be the ideal strategy because such designs will have a large $v$, thus limiting the rank of $\X$ and hence the ability to estimate second order effects. For example, suppose we have $m=8$ factor, $n=40$ runs, and we consider a two-factor interaction model which has $1+8+\binom{8}{2}=37$ model terms to estimate. All half designs will then have $n/2=20$ runs so $v=12$. A half design with $n_0=0$ and no replicates, will have $f=12$ fake factor degrees of freedom and $\text{rank}(\X)=n-f \leq 40-12=28$ so we can only estimate up to $28-9=19$ of the possible $28$ two-factor interaction effects. This problem worsens if we were to introduce center runs or replicates. We propose a new design and analysis strategy in Section 4 to address this issue.

\section{Design Construction Algorithms}\label{sec:alg}

The inefficiency of the ECI construction algorithm in \cite{Stallrich07082024} arose from most of the random initial designs failing to minimizing aliasing and only being able to guarantee a nonzero $g_X$ through forced replication. Both of these issues can be remedied by conditioning constructions on a foldover designs. Recall the main effect design variances for a $\D^\pm$ are half of the diagonal elements of $(\Hmat^T\Hmat)^{-1}$, the variance matrix for the main effect model without an intercept. This allows consideration of an $\Hmat$ with a constant column which would otherwise confound a main effect with the intercept.  The ECI criterion for a foldover design simplifies to minimizing 
\begin{align}
\frac{c(g_X)t_{\alpha/2,g_X} }{m} \sum_j \sqrt{v_j/2}\ ,\ \label{eqn:ECIfold}
\end{align}
where $g_X \geq f+p$ and $v_j$ is the $j$-th diagonal of $(\Hmat^T\Hmat)^{-1}$, being the half design's $j$-th design variance.

\subsection{General Algorithm}
We now present an efficient design search algorithm that finds a $\D^\pm$ that minimizes \eqref{eqn:ECIfold}. Theorem 1 tells us that a foldover design not only eliminates bias, but certain structures of $\Hmat$ can be enforced to generate pre-specified $f$ and $p$. The search algorithm uses the theorem to modify the coordinate exchange algorithm described in \cite{Stallrich07082024} by searching across half designs $\Hmat$ rather than arbitrary $\D$, for given $f$ and $p$. The new algorithm significantly increases efficiency over the original algorithm because (a) it only needs to search across $n/2$ runs instead of $n$ runs, (b) it guarantees zero aliasing, and (c) it can construct a design with a pre-specified $f$ and $p$.

The algorithm takes as inputs $n$, $m$, $n_0$, and a minimum number of rows of $\Hmat$ that are replicates of a non-center run, denoted by $R$. The choice of $n_0$ and $R$ will need to be chosen so that $\text{rank}(\Hmat)=m$ is possible. This ensures that the resulting $\D^\pm$ will have $f = v-n_0-R$ fake factor degrees of freedom and $p \geq \max(0,2n_0-1)+2R$ pure error degrees of freedom. Details of the algorithm may be found in \cite{Stallrich07082024}, but the basic idea is to partition the initial $\Hmat$ into unrestricted rows, $\Hmat_u$, and restricted rows, $\Hmat_r$. The restricted rows contain all $n_0$ center runs and $R$ rows that are required to be a replicate of a row in $\Hmat_u$. The search algorithm alternates between a coordinate exchange algorithm on $\Hmat_u$ and a row exchange on the non-center runs of $\Hmat_r$ with only the $\Hmat_u$ rows as candidates. When performing the coordinate exchange on some row in $\Hmat_u$, that row either is or is not replicated in $\Hmat_r$. If it is replicated in $\Hmat_r$, the coordinate exchange is also applied to the corresponding rows in $\Hmat_r$. The Supplementary Materials includes \texttt{R} code to perform the search algorithm.

It is possible for replicated rows to occur within $\Hmat_u$, particularly if such rows significantly decreased $c(g_X) \ t_{\alpha/2,g_X}$. However, $c(g_X) \ t_{\alpha/2,g_X}$ is bounded below by $z_{\alpha/2}$, creating a diminishing return in increasing $g_X$. We have found that $g_X=3$ or $4$ is usually sufficient for $\alpha=0.05$ (see the Supplementary Materials).  For large $n$ or odd $n$, we recommend an augmentation approach, described in Section~\ref{sec:FoldoverAug}.

When the proposed model does not include any quadratic effects, we recommend using the coordinates $\pm 1$ only, even if the factors are numeric. For each factor that is allowed a potential quadratic effect in the model, we require a row in $\Hmat_u$ that is unrestricted except for the coordinate for that factor, which is fixed at 0. This encourages each factor's quadratic effect to be estimable across submodels considered in the second stage analysis, although it is not guaranteed (e.g., design \textbf{R0.a05.n20} in the Supplementary Materials). This idea was motivated by the original DSD construction in \cite{jones2011class}, which assumed all factors had potential quadratic effects. We also recommend in this case that only the coordinates $\{0,\pm1\}$ be considered for such factors for faster design construction.

For a given $v$, we recommend Theorem 1 be used to determine all potential characterizations of $g_X$ and to use the algorithm to explore all scenarios. This ultimately comes down to specifying $n_0$ and $R$, which uniquely determines $f$ from \eqref{eqn:Thm1_f}. For example, if $v=1$ then there are only three possible choices: (A) $n_0=0$, $R=0$, $f=1$, (B) $n_0=1$, $R=0$, $f=0$, and (C) $n_0=0$, $R=1$, $f=0$. Case (A) has $p=0$ while cases (B) and (C) have $p=2$. However, if we performed case (A) in the search algorithm, the best design would likely include replication in $\Hmat_u$ because increasing $g_X$ from 1 to 2 will cause $t_{\alpha/2,g_X}$ to decrease significantly (e.g., $t_{0.025,1}=12.7062$ versus $t_{0.025,2}=4.3027$). Enumerating all the cases can be cumbersome for large $v$, but based on our previous recommendation of $g_X=3$ or $4$, the largest value to consider is $v=4$ unless there is concern about small signal to noise ratios.

To demonstrate, suppose we have $m=4$ factors with the first two factors having only two levels and the last two factors having potential quadratic effects. Then a valid initial half design with 6 runs and $R=2$ would be
\[
\Hmat=\left(\begin{array}{cccc}
+ & + & 0 & +\\
+ & + & - & 0\\
+ & - & + & -\\
+ & - & - & +\\
\hline
+ & + & 0 & +\\
+ & - & + & -\\
\end{array}\right)\ ,\
\]
where the horizontal line partitions the top $\Hmat_u$ and bottom $\Hmat_r$. The first row has a 0 in the third column and a 0 in the fourth column of the second row. The coordinate exchange algorithm will always skip these two coordinates, as well as the last two rows corresponding to $\Hmat_r$. The first row of $\Hmat_u$ is replicated in the first row of $\Hmat_r$, so a coordinate exchange on the first row would also be done on the first row in $\Hmat_r$. After all coordinate exchanges are considered in $\Hmat_u$, we then consider row exchanges in $\Hmat_r$ with those in the new $\Hmat_u$. This process continues until convergence. Throughout all of these exchanges, we maintain zero aliasing, $f=0$, and $p=4$. Additional replicated runs in $\Hmat_u$ will not occur with $R=2$ because that would cause $\text{rank}(\Hmat) < 4$. If we fixed $R=1$ instead, the search algorithm could produce designs with $f=1$ and $p=2$ or with $f=0$ and $p=4$. The latter case would occur if the algorithm decided to introduce a replicated row in $\Hmat_u$. If we fixed $R=0$ then $\Hmat_u=\Hmat$ and the search algorithm would only perform a coordinate exchange. It has the ability to produce designs with replicated runs, but also allows for the characterization not yet considered: $f=2$ and $p=0$. Such a design would have the largest $\text{rank}(\X)$ and so could perform better in the second stage analysis if the first stage analysis has sufficient power.

\subsection{Special Algorithm for Two Level Factors\label{Sec-constwolevel}}

When all factors have two-levels or when all quadratic effects are ignored, the matrix $(\Hmat^T\Hmat)^{-1}$ is equivalent to the variance matrix for an unbiased chemical balance weighing designs \citep[see][]{cheng2014optimal}. Therefore, if $\Hmat$ is $A$-optimal within the class of unbiased chemical balance weighing designs with $n/2$ runs, the corresponding $\D^\pm$ is also $A$-optimal for the main effect screening design framework with respect to all possible two-level foldover designs.
The $A$-optimal unbiased chemical balance weighing designs have been the subject of comprehensive investigation \citep{cheng1980optimality,cheng2014optimal, singh2015optimal}, encompassing both their theoretical optimality properties and systematic construction methodologies. We now categorize adaptations of the constructions of \cite{cheng2014optimal} based on the congruence condition $n/2 \equiv t (\text{mod } 4)$ for $t = 0,1,2,3$, and discuss possibilities for values of $f$ and $p$.
\begin{itemize}
    \item[\textbf{C0.}] \textbf{$n/2 \equiv 0 (\text{mod } 4)$}: The half design, $\Hmat^*_0$, that chooses any $m$ columns of a normalized Hadamard matrix of order $n/2$ is $A$-optimal \citep{cheng1980optimality, cheng2014optimal}.  In fact, such a design is universally optimal across all variance-based criteria since the design variances equal their minimum possible values. When $m > n/4 $, the rows of $\Hmat^*_0$ are distinct; otherwise, there may exist a set of columns with replicated rows \citep[see Construction (0) in][]{singh2015optimal}. Thus, too few columns may result in $p>0$ pure error degrees of freedom in $\D^{\pm}$. For example, the following half design with $m=5$ factors and $n/2 = 8$ runs results in $\D^{\pm}$ with $p= 0$:
\[
\Hmat^*_0=\left(\begin{array}{ccccc}
+ & + & + & + & +\\ \hline
+ & - & + & - & + \\ \hline
+ & + & - & - & +  \\ \hline
+ & - & - & + & + \\ \hline
+ & + & + & + & -\\ \hline
+ & - & + & - & - \\ \hline
+ & + & - & - & -  \\ \hline
+ & - & - & + & - \\ 
\end{array}\right)  \ .\
\]
However, if $ m=4$, and the last column of $\Hmat^*_0$ is deleted, the pure error degrees of freedom for $\Hmat^*_0$ and $\D^{\pm}$ are 4 and 8, respectively, since 1st and 5th row are replicates, 2nd and 4th row are replicates, and so on.

    \item[\textbf{C1.}] \textbf{$n/2 \equiv 1 (\text{mod } 4)$}: Construct the half-design $\Hmat^*_1$ by (a) first retaining any $m$ columns of a normalized Hadamard matrix of order $n/2-1$, and (b) adding \textit{any} row of $\pm 1$ to the reduced matrix obtained in (a). Following Theorem 2 in \cite{singh2015optimal}, this $\Hmat^*_1$ is $A$-optimal. This construction is an extension of \cite{cheng2014optimal}, which says to add a row of all $+1$'s in step (b).  If the added row of $\Hmat^*_1$ replicates one of the first $(n/2-1)$ rows, $p$ will be at least two. Otherwise $p$ depends on the presence/absence of replicates in the first $n/2-1$ runs. Furthermore, similar to case \textbf{C0}, if $m \ge n/4 $, then the rows of $\Hmat^*_1$ are distinct; otherwise, careful retaining of the columns may lead to further replicated rows in $\Hmat^*_1$, implying even larger $p$ for the foldover design $\D^{\pm}$. For example, $\Hmat^*_1$ for $m= 5$ and $n/2 = 9$ obtained by adding $(-\; -\; +\; + \; -)$ to $\Hmat^*_0$ presented in \textbf{C0} results in the foldover design $\D^{\pm}$ with $p = 2$. However, adding $(-\; +\; +\; + \; +)$  to $\Hmat^*_0$ results in the foldover design $\D^{\pm}$ with $p = 0$. 

   \item[\textbf{C2.}] \textbf{$n/2 \equiv 2 (\text{mod } 4)$, $v \geq 2$}:
    Construct the half-design $\Hmat^*_2$ by (a) first retaining any $m$ columns of a normalized Hadamard matrix of order $n/2-2$, and (b) adding a row of $+ 1$'s to the reduced matrix obtained in (a), and (c) add \textit{any} row consisting of $\pm 1$ such that the number of 1's and $-1$'s differ by at most 1. This construction is also an extension of \cite{cheng2014optimal}, which says to add a row with first (almost) half entries as $+1$'s and remaining $-1$'s in step (c). The $A$-optimality of such a $\Hmat^*_2$ again follows from the results in \cite{jacroux1983optimality} and \cite{singh2015optimal}. The first added row guarantees $p \geq 2$, since the first row of the normalized Hadamard matrix will be all $+1$'s. If the second added row is also a replicated row, $p \geq 4$. Again, careful column selection may introduce further replications in $\Hmat^*_2$, leading to an even larger $p$ in the foldover design $\D^{\pm}$. For example, $\Hmat^*_2$ for $m= 5$ and $n/2 = 10$ obtained by adding  a row of $+1$'s  and a $(-\; -\; +\; + \; -)$ to $\Hmat^*_0$ presented in \textbf{C0} results in the foldover design $\D^{\pm}$ with $p = 4$ since both added rows are each other's replicates. However, adding $(+\; -\; -\; - \; +)$  to $\Hmat^*_0$ along with a row of $+1$'s results in the foldover design $\D^{\pm}$ with $p = 2$ because of the replicate of the row of all $1$'s. 

    \item[\textbf{C3.}] \textbf{$n/2 \equiv 3 (\text{mod } 4)$}:  Construct the half-design $\Hmat^*_3$ by selecting any $m$ columns and deleting any row of a normalized Hadamard matrix of order $n/2+1$. This is the most difficult scenario and the $A$-optimality of such $\Hmat^*_3$ is only established for particular values of $m$ \citep[see][]{cheng1985optimal, singh2015optimal}. In addition, the $m$ should be small compared to $n/2+1$ to have $p \geq 2$, otherwise, $p=0$.
  
\end{itemize}
%

For faster design construction, we could consider only such constructions and then catalog designs based on their different $g_X$ characterizations, since all such designs will have the same $v_j$ values. The Supplementary Materials includes \texttt{R} code to perform the specialized search algorithm. A demonstration is given in Section~\ref{sec:Sim}



\section{Augmented Foldover Designs}\label{sec:FoldoverAug}

Conditioning our construction on foldover designs does come with some limitations, the most obvious one being that $n$ must be even. It would be foolish to remove a run just to make $n$ even. If all the factors are numeric, one could simply add a center run, similar to the DSD construction, but that is not feasible for two-level factors. Another limitation is that the foldover construction produces $g_X \geq v$ which implies $\text{rank}(\X) \leq n-v$. Since the lower bound of $c(g_X)t_{\alpha/2,g_X}$ is nearly attained when $g_X = 3$ or $4$, the marginal reduction in $c(g_X)t_{\alpha/2,g_X}$ for larger $v$ is negligible compared to the substantial loss in our ability to detect important second order effects caused by the reduced rank of $\X$. 

The issues encountered in situations with odd $n$ as well as $n$ with large $v$ can be fixed with a simple two-step construction algorithm. The algorithm first generates a foldover design targeting the first stage analysis with $v \leq 4$, followed by augmentation of that design with runs that target the second stage analysis. Based on the work by \cite{Stallrich02102023}, we choose the augmented runs by minimizing a Bayesian $A$-criterion with a large prior variance (e.g., $\tau^2 \geq 50$) for the second order terms to emphasize their estimation. Let $\X_0^T\X_0$ be the information matrix of the full model for the initial $\D^\pm_0$ and let $\X_A^T\X_A$ be the information matrix from the $n_A$ augmented runs. Then the Bayesian $A$-criterion for given $\tau^2$  minimizes 
\begin{align}
\text{tr}\left[ \left(\X_A^T\X_A + \X_0^T\X_0 + \frac{1}{\tau^2}\Kmat \right)^{-1} \right]\ ,\ \label{eqn:BayesA}
\end{align}
where $\Kmat$ is a diagonal matrix with $0$'s in the first $m+1$ diagonal elements and $1$'s in the remaining diagonal elements, corresponding to the second order terms. Note the optimization is done with respect to $\X_A^T\X_A$, as the other matrices are fixed. The criterion can be quickly optimized using a coordinate exchange algorithm, which we have included in the \texttt{R} code in the Supplementary Materials.

Performing the first stage analysis with all the runs of an augmented foldover design is generally not recommended because the design will likely not have zero aliasing. This would negate perhaps the most useful property of foldover designs. Fortunately, the structure of these augmented designs implies a partitioned analysis that leverages the structure of the design. We suggest the following steps for the analysis of augmented design.
\begin{enumerate}
    \item[(A)] Calculate $\hat{\sigma}^2_X$ from the full augmented foldover design.
    \item[(B)] Perform the first stage analysis on the largest subset of runs that have zero aliasing. This will likely be the initial $\D_0^\pm$.
    \item[(C)] Perform the preferred second stage analysis on the full augmented foldover design, conditioned on the active factors determined in step (B).
\end{enumerate}
For step (B) to be successful, the initial foldover design's run size should be large enough for the design to be sufficiently powered. The interpretation of the ECI criterion given after equation~\eqref{eqn:ECI} in Section~\ref{sec:Background} tells us that an ECI value of 1 implies a high probability of detecting active effects with a signal to noise ratio of 1. In our experience, this is adequate for most experiments. In general, practitioners should decide on a minimum signal to noise ratio that should be detected in the first stage, and ensure the ECI value of the resulting foldover design is less than or equal to that ratio. For Step (C), we recommend using the all subsets method from \cite{Stallrich07082024}. We demonstrate the construction and analysis of augmented foldover designs in Section~\ref{sec:Sim}.

\section{Simulation Studies}\label{sec:Sim}

This section applies the foldover design construction algorithms and the two stage analyses for two design scenarios, one with all two-level factors and one with all three-level factors. Each example compares different designs using either a full foldover or an augmented foldover design. The designs we generate algorithmically are based on 1000 random initial designs. The generated designs are compared in terms of the ECI values \eqref{eqn:ECIfold} of the foldover runs, average standard errors, and $g_X$ values. 

A simulation study is performed by varying the number of the active main effects, two-factor interactions, and quadratic terms (when applicable) under strong effect heredity. Active effects are generated from an exponential distribution with a rate parameter of $1$ and a positive offset term to set the minimum signal to noise ratio. Random signs are then applied to each of the effects. For each simulation setting, 500 responses are generated and analyzed for each competing design using the two-stage inference method. For augmented foldover designs, we employ the augmented analysis described in Section~\ref{sec:FoldoverAug} to ensure zero aliasing of the main effects. We report the true positive rates (TPRs) and false positive rates (FPRs) for the main effects, two-factor interactions, and quadratic effects.

\subsection{$n=16$, $m=5$ Two Level Factors}

With $m=5$ two-level factors, there are 15 possible effects in the second order model \eqref{eqn:QuadMod_Scalar}. For $n=16$ runs, a traditional approach would employ a half fraction that confounds the intercept with the five factor interaction. This is not a foldover design, but it can still estimate all main effects and two-factor interactions orthogonally. The drawback is the design does not have a pre-selection estimator of variance. A popular analysis approach is that by \cite{Lenth}, which generates a pseudo standard error estimator and margin of error term for a confidence interval. This analysis method will struggle if there are many active effects and/or if the magnitude of the active effects is small.

We generated foldover designs for $14$ runs and augmented them with two additional runs to target estimation of two-factor interactions. The half designs had $m+v=7$ runs so we considered the construction \textbf{C3} in Section~\ref{Sec-constwolevel}, which only gives designs with $f=g_X=2$. We then ran the algorithm described in Section~\ref{sec:alg} with initial settings $R \geq 1$ and $\alpha=0.05$, using 1,000 initial designs. This produced a competing design with $g_X=4$. 
To enforce $g_X=3$, we needed $R=1$, so we set $\alpha=0.75$ in the algorithm to diminish the utility in having an extra degree of freedom.  The three half designs and their statistical properties, as well as the augmented runs, denoted $\D_A$, are given in  Table~\ref{tab:Section5_1_Designs}.

\begin{table}[h]
    \centering
    \caption{Half designs, augmented runs, and properties of three competing designs for $m=5$ factors and $n=16$ runs. The ECI and the average (Avg) $\sqrt{v_j/2}$ values apply only to the foldover runs.}
    \begin{tabular}{r|c|c|c}
         Name      & \textbf{C3} & \textbf{R1.a05} & \textbf{R1.a75}\\ \hline
         $\Hmat$   & $\begin{pmatrix} 
                     + & + & + & + & +\\ \hline
                     + & - & + & - & +\\\hline
                     + & + & - & - & +\\ \hline
                     + & - & - & + & +\\ \hline
                     + & + & + & + & -\\ \hline
                     + & - & + & - & -\\ \hline
                     + & + & - & - & -
                     \end{pmatrix}$ & 
                     $\begin{pmatrix} 
                     - & - & + & - & +\\
                     - & - & + & - & +\\ \hline
                     + & + & - & - & +\\
                     - & - & + & + & -\\ \hline
                     + & - & + & - & -\\ \hline
                     - & - & - & - & -\\ \hline
                     - & + & + & - & -\\ 
                     
                     \end{pmatrix}$ &
                     $\begin{pmatrix} 
                     + & + & - & + & +\\
                     + & + & - & + & +\\ \hline
                     + & - & + & + & -\\ \hline
                     - & + & - & + & -\\ \hline
                     - & + & + & + & +\\ \hline
                     - & - & - & + & +\\ \hline
                     - & - & + & + & +\\
                     \end{pmatrix}$\\ \hline
         $\D_A$ & $\begin{pmatrix} 
         + & - & - & + & -\\
         + & - & + & + & -\end{pmatrix}$ & $\begin{pmatrix} 
         +  &  +  & - &  +  &  +\\
         +  &  +  & - &  -  & -\end{pmatrix}$ & $\begin{pmatrix} 
         + &  -  &  + & -  & -\\
         + &  + &  -  & - &  +\end{pmatrix}$\\ \hline
         Foldover ECI       & $1.101$ & $0.777$ & $0.865$\\
         Avg $\sqrt{v_j/2}$ & $0.289$ & $0.298$ & $0.295$\\
         $(f,p,\ell_X,g_X)$ & $(2,0,2,2)$ & $(0,4,0,4)$ & $(1,2,1,3)$
    \end{tabular}
\label{tab:Section5_1_Designs}
\end{table}

A simulation study compared the analysis proposed in Section~\ref{sec:FoldoverAug} for the three augmented designs from Table~\ref{tab:Section5_1_Designs}, to the Lenth analysis of the half fraction under different effect distributions. We considered six different scenarios corresponding to different settings of active effects and effect offset values. The simulation scenarios and the results are shown in Table~\ref{tab:Section5_2_Analysis}. All four designs had high TPRs and low FPRs for sparse models with large signal to noise ratios. The half fraction design was better than the other designs for the sparsest model and small signal to noise ratio, but quickly became worse as the model became less sparse. Its performance is very troubling for the models with a total of 10 active effects, having TPRs less than $0.10$. Among the three augmented foldover designs we generated, design \textbf{R1.a05} performed the best due to its higher TPRs for main effects, although this does come with some reduction in TPR of the interaction effects in some scenarios.

\begin{table}[ht]
    \centering
     \caption{Simulation study scenarios and results for three $m=5$, $n=16$ designs shown in Table~\ref{tab:Section5_1_Designs} and $2^{5-1}$ half fraction design. Main effect FPRs are blank for the case of 5 active main effects because all factors are active.}
    \begin{tabular}{ccr|cc|cc}
    \multicolumn{2}{c}{\# Active (SN)} &  & \multicolumn{2}{c|}{Main} & \multicolumn{2}{c}{2FIs}\\
    Main & 2FIs & Design & TPR & FPR 
        & TPR & FPR\\ \hline
    3 (2) & 2 (1) & \textbf{C3 }& $0.977$ & $0.054$ & $0.943$ & $0.035$\\
      &   &    \textbf{R1.a05} & $1.000$ & $0.049$ & $0.983$ & $0.035$\\
      &   &    \textbf{R1.a75} & $0.999$ & $0.056$ & $0.985$ & $0.032$\\
      &   &    Half Fraction & $1.000$ & $0.020$ & $0.957$ & $0.020$\\ \hline

    3 (0.75) & 2 (0.5) & \textbf{C3} & $0.700$ & $0.048$ & $0.508$ & $0.032$\\
      &   &    \textbf{R1.a05} & $0.853$ & $0.040$ &  $0.665$ &  $0.025$\\
      &   &    \textbf{R1.a75} & $0.800$ & $0.038$ & $0.601$ & $0.025$\\
      &   &    Half Fraction & $0.820$ & $0.013$ & $0.957$ & $0.020$\\ \hline

    4 (2) & 3 (1) & \textbf{C3} & $0.972$ & $0.051$ & $0.849$ & $0.089$\\
      &   &    \textbf{R1.a05} & $0.999$ & $0.048$ & $0.741$ & $0.119$\\
      &   &    \textbf{R1.a75} & $0.999$ & $0.047$ & $0.847$ & $0.102$\\
      &   &    Half Fraction & $0.988$ & $0.012$ & $0.887$ & $0.007$\\ \hline

    4 (0.75) & 3 (0.5) & \textbf{C3} & $0.690$ & $0.053$ & $0.441$ & $0.074$\\
      &   &    \textbf{R1.a05} & $0.856$ & $0.045$ &  $0.536$ &  $0.103$\\
      &   &    \textbf{R1.a75} & $0.814$ & $0.048$ & $0.530$ & $0.098$\\
      &   &    Half Fraction & $0.641$ & $0.002$ & $0.531$ & $0.002$\\ \hline

    5 (2) & 5 (1) & \textbf{C3} & $0.976$ & $-$ & $0.688$ & $0.288$\\
      &   &    \textbf{R1.a05} & $1.000$ & $-$ & $0.466$ & $0.304$\\
      &   &    \textbf{R1.a75} & $0.999$ & $-$ & $0.594$ & $0.309$\\
      &   &    Half Fraction & $0.025$ & $-$ & $0.008$ & $0.007$\\ \hline

    5 (0.75) & 5 (0.5) & \textbf{C3} & $0.710$ & $-$ & $0.395$ & $0.204$\\
      &   &    \textbf{R1.a05} & $0.869$ & $-$ &  $0.377$ & $0.254$\\
      &   &    \textbf{R1.a75} & $0.820$ & $-$ & $0.411$ & $0.265$\\
      &   &    Half Fraction & $0.095$ & $-$ & $0.065$ & $0.002$\\ \hline
    \end{tabular}
    \label{tab:Section5_1_Analysis}
\end{table}

\subsection{$n=24$, $m=7$ Three Level Factors\label{sec-exthree}}

\cite{Stallrich07082024} constructed a screening design, denoted here as \textbf{SM.n22}, with $n=24$ runs and $m=7$ three level factors using the less efficient ECI criterion. Among their constructed designs with an ECI value less than 1, they chose the one they felt did best in terms of estimating the second order effects. Among the 24 runs, 22 of them were foldover pairs, and had an augmented foldover structure with $f=4$ fake factor degrees of freedom. We used the more efficient algorithm from Sections~\ref{sec:alg} and \ref{sec:FoldoverAug} to create six competing designs for the comparison against \textbf{SM.n22}. The first three designs were full foldovers with $(R,n_0,\alpha)$ set to $(0,0,0.05)$, $(0,0,0.75)$ and $(1,1,0.05)$. The last three designs were partial foldovers of $20$ runs with the same $(R,n_0,\alpha)$ settings as the full foldovers, and $n_A=4$ augmented runs. The half designs and augmented runs may be found in the Supplementary Materials, and the design properties are summarized in Table~\ref{tab:Section5_2_Designs}.

\begin{table}[h]
    \centering
    \caption{Design properties of the seven competing designs for $n=24$ runs and $m=7$ three-level factors. The ECI and Avg $\sqrt{v_j/2}$ values apply only to the foldover runs.}
    \resizebox{\columnwidth}{!}{
    \begin{tabular}{r|ccc|ccc|c}
         Name& \textbf{ADSD.n24} & \textbf{R0.a05.n24} & \textbf{R1.n01.a05.n24} & \textbf{R0.a75.n20} & \textbf{R0.a05.n20} & \textbf{R1.n01.a05.n20} & \textbf{SM.n22}\\ \hline
         $n_A$& 0 & 0 & 0 & $4$ & $4$ & $4$ & $2$\\
         ECI &  $0.521$ & $0.511$ & $0.533$ & $0.691$ & $0.631$ & $0.672$ & $0.729$\\
         Avg $\sqrt{v_j/2}$ & $0.213$ & $0.224$ & $0.239$ & $0.236$ & $0.258$ & $0.257$ & $0.279$\\
         $(f,p,\ell_X,g_X)$ & $(5,0,5,5)$ & $(3,4,3,7)$ & $(1,7,1,8)$ & $(3,0,3,3)$ & $(1,4,1,5)$ & $(1,3,1,4)$ & $(4,0,4,4)$
    \end{tabular}
    }
\label{tab:Section5_2_Designs}
\end{table}
Design \textbf{ADSD.n24} referenced in Table~\ref{tab:Section5_2_Designs} is generated using $(R,n_0,\alpha)=(0,0,0.75)$, and is the augmented DSD mentioned in \cite{Stallrich07082024} with the center run removed. Note that the less efficient algorithm of \cite{Stallrich07082024} was never able to find this design despite using 5000 initial designs. Design \textbf{R0.a05.n24} has a smaller ECI value than \textbf{ADSD.n24}, meaning it will have higher power for detecting main effects, primarily due to it having $g_X=7$ versus $g_X=5$. Design \textbf{SM.n22} has the worst ECI value, primarily because it was not chosen to minimize the ECI criterion. Rather, it was chosen among all designs with an ECI value less than 1, such that it has the best estimation of second order effects.

A simulation study compared the analysis of Section~\ref{sec:FoldoverAug} of the seven designs from Table~\ref{tab:Section5_2_Designs} under three different effect distributions. The simulation scenarios and the results are shown in Table~\ref{tab:Section5_2_Analysis}.  The scenarios focus on larger models with fixed signal to noise ratios for the main effects, two-factor interactions, and quadratic effects to be $1.5$, $2.5$, and $2.5$, respectively, following \cite{Stallrich07082024}. All designs had high TPRs and low FPRs for the main effects, which is expected given the ECI values from Table~\ref{tab:Section5_2_Designs} are all well below $1.5$. This was also true of all second order effects for the smallest model considered. \textbf{ADSD.n24} had the highest TPR for the two-factor interactions for the first two scenarios, but surprisingly \textbf{R0.a75.n20} had a higher TPR ($0.908$ vs $0.826$) for the last scenario with the largest model. \textbf{R0.a75.n20} also had a higher TPR than \textbf{ADSD.n24} for the quadratic effects (0.635 vs 0.539), with \textbf{SM.n22} having the highest TPR (0.712). However, the increase in quadratic effect TPR for \textbf{SM.n22} comes with a decrease in TPR for interactions ($0.777$). These results demonstrate the value of the augmented design and analysis approach.

\begin{table}[ht]
    \centering
     \caption{Simulation study cases and results for seven $k=6$, $n=24$ designs. Following \cite{Stallrich07082024}, we fixed the signal to noise ratios for the main effects, two-factor interactions, and quadratic effects to be $1.5$, $2.5$, and $2.5$, respectively.}
    \begin{tabular}{cccr|cc|cc|cc}
    \multicolumn{3}{c}{\# Active} &  & \multicolumn{2}{c|}{Main} & \multicolumn{2}{c|}{2FIs} & \multicolumn{2}{c}{Quad}\\
    Main & 2FIs & Quad & Design & TPR & FPR 
        & TPR & FPR & TPR & FPR \\ \hline
    3 & 2 & 2 & \textbf{ADSD.n24} & $1.000$ & $0.048$ &  $0.999$  & $0.015$  &  $0.961$  & $0.034$\\
      &   &   & \textbf{R0.a05.n24} &  $1.000$ & $0.053$ &  $0.985$  & $0.013$  &  $0.945$  & $0.026$\\
      &   &   & \textbf{R1.n01.a05.n24} &  $1.000$ & $0.051$ &  $0.983$  & $0.011$  &  $0.972$  & $0.032$\\
      &   &   & \textbf{R0.a75.n20} &  $0.995$ & $0.054$ &  $0.989$  & $0.008$  &  $0.909$  & $0.013$\\
      &   &   & \textbf{R0.a05.n20} &  $0.999$ & $0.056$ &  $0.990$  & $0.010$  &  $0.905$  & $0.027$\\
      &   &   & \textbf{R1.n01.a05.n20} &  $0.999$ & $0.048$ &  $0.993$  & $0.006$  &  $0.927$  & $0.013$\\
      &   &   & \textbf{SM.n22} &  $0.994$ & $0.036$ &  $0.980$  & $0.013$  &  $0.977$  & $0.030$\\ \hline
    5 & 3 & 1 & \textbf{ADSD.n24} & $1.000$ &  $0.047$ &  $0.988$ &  $0.037$ &    $0.878$ &   $0.057$ \\
      &   &   & \textbf{R0.a05.n24} &  $1.000$ &  $0.043$ &  $0.913$ &  $0.035$ &    $0.792$ &   $0.059$\\
      &   &   & \textbf{R1.n01.a05.n24} &  $1.000$ &  $0.039$ &  $0.807$ &  $0.038$ &    $0.754$ &   $0.086$\\
      &   &   & \textbf{R0.a75.n20} &  $0.996$ &  $0.044$ &  $0.978$ &  $0.024$ &    $0.888$ &   $0.036$\\
      &   &   & \textbf{R0.a05.n20} &  $0.999$ &  $0.038$ &  $0.955$ &  $0.025$ &    $0.804$ &   $0.054$ \\
      &   &   & \textbf{R1.n01.a05.n20} &  $0.997$ &  $0.051$ &  $0.957$ &  $0.027$ &    $0.870$ &   $0.052$ \\
      &   &   & \textbf{SM.n22} &  $0.996$ &  $0.049$ &  $0.941$ &  $0.045$ &    $0.926$ &   $0.070$\\ \hline
    5 & 4 & 3 & \textbf{ADSD.n24} & $1.000$ &  $0.050$ &  $0.826$ &  $0.088$ &   $0.539$ &   $0.118$\\
      &   &   & \textbf{R0.a05.n24} &  $1.000$ &  $0.067$ &  $0.584$ &  $0.106$ &   $0.342$ &   $0.159$\\
      &   &   & \textbf{R1.n01.a05.n24} &  $1.000$ &  $0.041$ &  $0.452$ &  $0.100$ &   $0.337$ &   $0.171$\\
      &   &   & \textbf{R0.a75.n20} & $0.998$ &  $0.042$ &  $0.908$ &  $0.049$ &   $0.635$ &   $0.061$\\
      &   &   & \textbf{R0.a05.n20} & $0.999$ &  $0.041$ &  $0.715$ &  $0.081$ &   $0.483$ &   $0.129$\\
      &   &   & \textbf{R1.n01.a05.n20} & $0.998$ &  $0.039$ &  $0.766$ &  $0.071$ &   $0.506$ &   $0.010$\\
      &   &   & \textbf{SM.n22} & $0.996$ &  $0.055$ &  $0.777$ &  $0.089$ &   $0.712$ &   $0.104$\\ \hline
    \end{tabular}
    \label{tab:Section5_2_Analysis}
\end{table}

\newpage
\section{Ethylene Experiment\label{sec-expt}}

Prior to each run, a fresh salt sample consisting of 60 mol\% Li\textsubscript{2}CO\textsubscript{3} - 20 mol\% Na\textsubscript{2}CO\textsubscript{3} - 20 mol\% K\textsubscript{2}CO\textsubscript{3} was prepared via physical mixing from anhydrous Li\textsubscript{2}CO\textsubscript{3} (Sigma Aldrich, $>$ 99.0\% purity), Na\textsubscript{2}CO\textsubscript{3} (Sigma Aldrich, $>$ 99.5\% purity), and K\textsubscript{2}CO\textsubscript{3} (Sigma Aldrich, $>$ 99\% purity) precursors. Each prepared sample was transferred to a clean, closed-end alumina (99.8\% Al\textsubscript{2}O\textsubscript{3}) reaction vessel and fitted to a Swagelok reactor assembly. Suspended in a high-temperature vertical furnace, each sample was then heated to 800°C (ramp rate = 20°C/min) under a CO\textsubscript{2} atmosphere to mitigate carbonate decomposition. Alicat mass flow controllers metered the feed gases to electrically-actuated Emerson ASCO solenoid valves, which provided control of gas flow into the reactor. Inlet gases entered the reactor through an open-ended alumina injection tube, which was submerged into the molten salt upon melting. A 1/16” sheathed thermocouple and bubbled the gas phase through the molten mixture.  

Each MM-ODH cycle consisted of an ethane oxidation step and a molten carbonate regeneration step separated by two inert gas purges. During carbonate regeneration, CO\textsubscript{2} diluted in argon (Ar) re-carbonated the salt prior to starting the next cycle. 37.5 SCCM of Ar purged the system between each step. Each MM-ODH cycle operated isothermally and isobarically at 800°C and \~1 bar, respectively. An in-line MKS Cirrus III quadrupole mass spectrometer carried out residual gas analysis (RGA) of the reactor effluent to determine real-time gas compositions. Gas samples for GC analysis were collected in Tedlar gas bags, with each collection spanning the entirety of the oxidation or regeneration step and 5 minutes of the following purge. These samples were analyzed using an Agilent 7890B gas chromatography unit equipped with a flame ionization detector (FID) and two thermal conductivity detectors (TCD). OpenLab EZChrom software executed residence-time-based peak identification based on Agilent gas standard calibration.

Seven responses were collected but we will focus on the percent of ethylene. Only two levels were considered for each factor and the budget allowed $n=20$ runs. The design was generated using the less efficient ECI algorithm from \cite{Stallrich07082024}, resulting in a full foldover with $f=1$, $p=2$, and $g_X=3$. All of the main effects had a constant absolute correlation of $0.2$ and an average design standard error of $0.270$. The ECI value is $0.791$ with $\alpha=0.05$ so the design has a power of roughly $0.95$ for main effects having a signal to noise ratio of $0.791$, despite the columns not being mutually orthogonal. Similar to design \textbf{SM.n22} from Section~\ref{sec-exthree}, the design was chosen among all designs with an ECI value less than 1 due to its better estimation of two-factor interactions. The data are shown in Table~\ref{tab:EthyleneData}, with the runs arranged as replicates and foldover pairs.

\begin{table}[ht]
    \centering
     \caption{Data for the Ethylene experiment. Each column represents a run and its corresponding response. The runs are arranged as foldover pairs.}
    \resizebox{\columnwidth}{!}{%
    \begin{tabular}{c|cc|cc|cc|cc|cc|cccc|cc|cc|cc|} 
    \hline
1   & $+$ & $-$ &  $+$ & $-$ &  $+$ & $-$ &  $+$ & $-$ &  $+$ & $-$ & $-$ & $-$ &  $+$ &   $+$ & $-$ &  $+$ &   $+$ & $-$ &  $+$ & $-$\\
2   & $-$ &  $+$ & $-$ &  $+$ &  $+$ & $-$ & $-$ &  $+$ &  $+$ & $-$ & $-$ & $-$ &  $+$ &   $+$ &  $+$ & $-$ &   $+$ & $-$ &  $+$ & $-$\\
3   & $+$ & $-$ & $-$ &  $+$ & $-$ &  $+$ &  $+$ & $-$ &  $+$ & $-$ &  $+$ &  $+$ & $-$ &  $-$ &  $+$ & $-$ &  $-$ &  $+$ &  $+$ & $-$ \\
4   & $+$ & $-$ & $-$ &  $+$ & $-$ &  $+$ &  $+$ & $-$ & $-$ &  $+$ &  $+$ &  $+$ & $-$ &  $-$ & $-$ &  $+$ &   $+$ & $-$ & $-$ &  $+$ \\
5   & $+$ & $-$ & $-$ &  $+$ &  $+$ & $-$ &  $+$ & $-$ &  $+$ & $-$ & $-$ & $-$ &  $+$ &   $+$ &  $+$ & $-$ &  $-$ &  $+$ & $-$ &  $+$ \\
6   & $-$ &  $+$ & $-$ &  $+$ &  $+$ & $-$ &  $+$ & $-$ & $-$ &  $+$ & $-$ & $-$ &  $+$ &   $+$ & $-$ &  $+$ &  $-$ &  $+$ &  $+$ & $-$ \\
7   & $-$ &  $+$ & $-$ &  $+$ &  $+$ & $-$ & $-$ &  $+$ &  $+$ & $-$ &  $+$ &  $+$ & $-$ &  $-$ & $-$ &  $+$ &  $-$ &  $+$ &  $+$ & $-$ \\
8   & $-$ &  $+$ &  $+$ & $-$ &  $+$ & $-$ &  $+$ & $-$ & $-$ &  $+$ &  $+$ &  $+$ & $-$ &  $-$ &  $+$ & $-$ &   $+$ & $-$ & $-$ &  $+$ \\ \hline
$y$ &  $0.30$ & $0.65$ & $0.41$  & $0.53$ & $0.62$ & $0.42$  & $0.28$ & $0.65$ & $0.64$ & $0.33$ & $0.39$ & $0.35$ & $0.57$   &$0.60$ & $0.66$ & $0.25$   &$0.50$ & $0.41$ & $0.59$ & $0.35$
    \end{tabular}
}
    \label{tab:EthyleneData}
\end{table}

The results of the first stage analysis are shown in Table~\ref{tab:EthyleneFirstStage}, which concluded factors 1, 2, and 4 as statistically significant with $\alpha=0.05$. We considered the all-subsets analysis with strong effect heredity for this set of factors. Table~\ref{tab:EthyleneSecondStage} shows the different interaction models considered in this analysis and their mBIC values (see \cite{Stallrich07082024}). Based on these results, we included the two-factor interaction between factors 1 and 4 in the final model. This simple model fitted the data well with an $R^2=0.967$.  If $\alpha=0.10$ had been used in the first stage, factor 6 would have also been included and the all-subsets analysis would again include the two-factor interaction between factors 1 and 4 in the final model. This is another competitive model, with an mBIC value of $29.204$ and $R^2=0.982$.

\begin{table}[ht]
    \centering
     \caption{First stage analysis results for the Ethylene experiment. Inference is based on the pre-selection estimate $\hat{\sigma}_X=0.024$ having 3 degrees of freedom.}
    \begin{tabular}{c|cccccc} 
    
  Factor & Estimates & StdErrors      & $T$  & $p$-value & Lower 95\% CI & Upper 95\% CI\\ \hline
     1  &  $-0.025$  &    $0.006$ & $-4.161$ & $0.025$ & $-0.045$  & $-0.006$ \\
      2 &    $\phantom{-}0.106$ &    $0.007$ & $14.907$ & $0.001$ &  $\phantom{-}0.083$  & $\phantom{-}0.128$ \\
      3 &    $\phantom{-}0.008 $&    $0.007$ &  $\phantom{-}1.113$ & $0.347$ & $-0.014$  &  $\phantom{-}0.029$ \\
      4 &   $-0.053$ &    $0.007$ & $-7.498$ & $0.005$ & $-0.076$  & $-0.031$ \\
      5 &   $-0.004$ &    $0.007$ & $-0.619$ & $0.580$ & $-0.025 $ & $\phantom{-}0.017$ \\
      6 &   $-0.015$ &    $0.006$ & $-2.460$ & $0.091$ & $-0.035 $ & $\phantom{-}0.004$ \\
      7 &   $-0.003$ &    $0.007$ & $-0.371$ & $0.735$ & $-0.024$  & $\phantom{-}0.019$ \\
      8 &    $\phantom{-}0.003$ &    $0.006$ &  $\phantom{-}0.462$  & $0.675$ & $-0.017$  & $\phantom{-}0.022$
    \end{tabular}
    \label{tab:EthyleneFirstStage}
\end{table}

\begin{table}[ht]
    \centering
     \caption{Second stage analysis results for the Ethylene experiment with factors 1, 2, and 4. An X indicates the effect was included in the model. Adding $d_1d_4$ decreased mBIC from 36.590 to 36.077.}
    \begin{tabular}{ccc|c} 
  $d_1d_2$ & $d_1d_4$ & $d_2d_4$  & mBIC\\ \hline
           &          &           &  36.590    \\ \hline
       X   &          &           &  37.867    \\ \hline
           &      X   &           &  36.077    \\ \hline
           &          &      X    &  38.270    \\ \hline
       X   &     X    &           &  39.000    \\ \hline
       X   &          &      X    &  39.825    \\ \hline
           &     X    &      X    &  38.149    \\ \hline
       X   &    X     &      X    &  41.097    \\ \hline
    \end{tabular}
    \label{tab:EthyleneSecondStage}
\end{table}


\newpage
\section{Discussion\label{sec-disc}}

Response surface methodology heavily relies on identifying small-run designs that balance factor screening and identifying a second order model to perform optimization. Following \cite{gilmour2012optimum,jones2017effective} and \cite{Stallrich07082024}, we strongly believe that designs having a pre-selection estimator for variance, $\hat{\sigma}^2_X$, can perform as well or better than traditional designs, with significant improvement when the common effect principles are violated. This paper shows that the foldover technique provides a rich class of designs that not only have zero aliasing between main effects and second order terms, but also generate degrees of freedom for $\hat{\sigma}^2_X$. Theorem 1 provides a characterization of the fake factor and pure error degrees of freedom for foldover designs. We also provide a fast construction algorithm and direct construction method to find a foldover design that minimizes the Expected Confidence Interval (ECI) criterion from \cite{Stallrich07082024} for some confidence level, $1-\alpha$. The ECI criterion is attractive to practitioners because it may be interpreted as the approximate minimum effect size of the main effects for which the power is at least $1-\alpha$. The algorithm is faster than that given in \cite{Stallrich07082024} because it needs to only optimize the half design, having half the number of runs as the foldover design.

An important consequence of Theorem 1 is that foldover designs with a large number of runs should be avoided because they will have an unnecessarily large number of degrees of freedom or $\hat{\sigma}^2_X$. Such designs will be severely limited in their ability to identify important second order effects. We propose an augmented design construction to add runs to a smaller foldover design to improve estimation of the second order terms. This is only recommended if the initial foldover design is already well-powered for the stage 1 analysis, which can be assessed using the ECI criterion value's interpretation. The augmented runs are chosen based on the Bayesian $A$-criterion with large prior variance on the second order terms, thereby focusing the augmented runs on improving estimation of these terms. This augmentation strategy can also be used to create designs when there are an odd number of runs, since foldover designs, by definition, must have an even number of runs. To maintain zero aliasing in the stage 1 analysis, we recommend analyzing only the foldover design's runs. The stage 2 analysis uses all of the runs in the design.

The augmented foldover construction and analysis presented in Section~\ref{sec:FoldoverAug} could be employed under a sequential design strategy in which an experiment using only the foldover runs is first performed and analyzed. One could then hone the second order model employed in \eqref{eqn:BayesA} to those second order effects that are consistent with effect heredity of the active factors. However, it would be generally recommended that a block effect be introduced in the model, requiring a small modification to \eqref{eqn:BayesA}.

The design construction algorithm in this paper primarily focuses on main effect screening via the ECI criterion. Practitioners may find that the best design found is overpowered for their expected effect size. Following \cite{Stallrich07082024}, all designs having an ECI value less than or equal to the expected effect size could be retained and evaluated using a secondary criterion. One example is the reduced lack of fit criterion from \cite{Stallrich02102023} which assesses a design's ability to perform model selection in the second stage.

\bibliographystyle{asa} 
\bibliography{main.bib}

\begin{thebibliography}{31}
\newcommand{\enquote}[1]{``#1''}
\expandafter\ifx\csname natexlab\endcsname\relax\def\natexlab#1{#1}\fi

\bibitem[{Ares and Goos(2020)}]{OMARS}
Ares, J.~N. and Goos, P. (2020), \enquote{Enumeration and Multicriteria
  Selection of Orthogonal Minimally Aliased Response Surface Designs,}
  \textit{Technometrics}, 62, 21--36.

\bibitem[{Cheng(1980)}]{cheng1980optimality}
Cheng, C.-S. (1980), \enquote{Optimality of some weighing and $2^n$ fractional
  factorial designs,} \textit{The Annals of Statistics}, 436--446.

\bibitem[{Cheng(2014)}]{cheng2014optimal}
--- (2014), \enquote{Optimal biased weighing designs and two-level main-effect
  plans,} \textit{Journal of Statistical Theory and Practice}, 8, 83--99.

\bibitem[{Cheng et~al.(1985)Cheng, Masaro, and Wong}]{cheng1985optimal}
Cheng, C.-S., Masaro, J.~C., and Wong, C.~S. (1985), \enquote{Optimal weighing
  designs,} \textit{SIAM Journal on Algebraic Discrete Methods}, 6, 259--267.

\bibitem[{Cheng and Tang(2005)}]{Min_G_ab}
Cheng, C.~S. and Tang, B. (2005), \enquote{A {G}eneral {T}heory of {M}inimum
  {A}berration and {I}ts {A}pplications,} \textit{The Annals of Statistics},
  33, 944--958.

\bibitem[{Cox and Reid(2000)}]{cox2000theory}
Cox, D.~R. and Reid, N. (2000), \textit{The {T}heory of the {D}esign of
  {E}xperiments}, Chapman and Hall/CRC.

\bibitem[{Deng and Tang(1999)}]{GMA}
Deng, L.~Y. and Tang, B. (1999), \enquote{Generalized resolution and minimum
  aberration criteria for {P}lackett-{B}urman and other nonregular factorial
  designs,} \textit{Statistica Sinica}, 9, 1071--1082.

\bibitem[{Diamond(1995)}]{Diamond1995}
Diamond, N.~T. (1995), \enquote{Some Properties of a Foldover Design,}
  \textit{Australian Journal of Statistics}, 37, 345--352.

\bibitem[{DuMouchel and Jones(1994)}]{dumouchel_jones1994}
DuMouchel, W. and Jones, B. (1994), \enquote{A Simple {B}ayesian Modification
  of {D}-Optimal Designs to Reduce Dependence on an Assumed Model,}
  \textit{Technometrics}, 36, 37--47.

\bibitem[{Errore et~al.(2017)Errore, Jones, Li, and
  Nachtsheim}]{Errore02012017}
Errore, A., Jones, B., Li, W., and Nachtsheim, C.~J. (2017), \enquote{Benefits
  and Fast Construction of Efficient Two-Level Foldover Designs,}
  \textit{Technometrics}, 59, 48--57.

\bibitem[{Gilmour and Trinca(2012)}]{gilmour2012optimum}
Gilmour, S.~G. and Trinca, L.~A. (2012), \enquote{Optimum design of experiments
  for statistical inference,} \textit{Journal of the Royal Statistical Society:
  Series C (Applied Statistics)}, 61, 345--401.

\bibitem[{Jacroux et~al.(1983)Jacroux, Wong, and
  Masaro}]{jacroux1983optimality}
Jacroux, M., Wong, C.~S., and Masaro, J.~C. (1983), \enquote{On the optimality
  of chemical balance weighing designs,} \textit{Journal of Statistical
  Planning and Inference}, 8, 231--240.

\bibitem[{Jones et~al.(2020)Jones, Hunter, and Montgomery}]{jones2020partial}
Jones, B., Hunter, J.~S., and Montgomery, D.~C. (2020), \enquote{Partial
  {R}eplication of {D}efinitive {S}creening {D}esigns,} \textit{Quality
  Engineering}, 32, 4--9.

\bibitem[{Jones and Nachtsheim(2011{\natexlab{a}})}]{jones2011class}
Jones, B. and Nachtsheim, C.~J. (2011{\natexlab{a}}), \enquote{A {C}lass of
  {T}hree-level {D}esigns for {D}efinitive {S}creening in the {P}resence of
  {S}econd-{O}rder {E}ffects,} \textit{Journal of Quality Technology}, 43,
  1--15.

\bibitem[{Jones and Nachtsheim(2011{\natexlab{b}})}]{Jones_2011}
--- (2011{\natexlab{b}}), \enquote{Efficient {D}esigns with {M}inimal
  {A}liasing,} \textit{Technometrics}, 53, 62--71.

\bibitem[{Jones and Nachtsheim(2017)}]{jones2017effective}
--- (2017), \enquote{Effective {D}esign-{B}ased {M}odel {S}election for
  {D}efinitive {S}creening {D}esigns,} \textit{Technometrics}, 59, 319--329.

\bibitem[{Lenth(1989)}]{Lenth}
Lenth, R. (1989), \enquote{Quick and easy analysis of unreplicated
  fractionals,} \textit{Technometrics}, 31, 469--473.

\bibitem[{Leonard and Edwards(2017)}]{leonard2017bayesian}
Leonard, R.~D. and Edwards, D.~J. (2017), \enquote{Bayesian {D}-optimal
  screening experiments with partial replication,} \textit{Computational
  Statistics \& Data Analysis}, 115, 79--90.

\bibitem[{Li and Lin(2003)}]{Li01052003}
Li, W. and Lin, D. K.~J. (2003), \enquote{Optimal Foldover Plans for Two-Level
  Fractional Factorial Designs,} \textit{Technometrics}, 45, 142--149.

\bibitem[{Lin et~al.(2008)Lin, Miller, and Sitter}]{LIN20083107}
Lin, C.~D., Miller, A., and Sitter, R. (2008), \enquote{Folded over
  non-orthogonal designs,} \textit{Journal of Statistical Planning and
  Inference}, 138, 3107--3124.

\bibitem[{Mead et~al.(2012)Mead, Gilmour, and Mead}]{mead2012statistical}
Mead, R., Gilmour, S.~G., and Mead, A. (2012), \textit{Statistical {P}rinciples
  for the {D}esign of {E}xperiments: {A}pplications to {R}eal {E}xperiments},
  vol.~36, Cambridge University Press.

\bibitem[{Miller and Sitter(2001)}]{Miller01022001}
Miller, A. and Sitter, R.~R. (2001), \enquote{Using the Folded-Over 12-Run
  Plackett—Burman Design to Consider Interactions,} \textit{Technometrics},
  43, 44--55.

\bibitem[{Miller and Sitter(2005)}]{Miller01112005}
--- (2005), \enquote{Using Folded-Over Nonorthogonal Designs,}
  \textit{Technometrics}, 47, 502--513.

\bibitem[{Montgomery(2019)}]{Montgomery2019}
Montgomery, D.~C. (2019), \textit{Design and {A}nalysis of {E}xperiments}, New
  York: Wiley, 10th ed.

\bibitem[{Nguyen et~al.(2020)Nguyen, Pham, and Mai}]{Nguyen02012020}
Nguyen, N.-K., Pham, T.-D., and Mai, P.~V. (2020), \enquote{Constructing
  D-Efficient Mixed-Level Foldover Designs Using Hadamard Matrices,}
  \textit{Technometrics}, 62, 48--56.

\bibitem[{Singh et~al.(2015)Singh, Chai, and Das}]{singh2015optimal}
Singh, R., Chai, F.-S., and Das, A. (2015), \enquote{Optimal two-level choice
  designs for any number of choice sets,} \textit{Biometrika}, 102, 967--973.

\bibitem[{Stallrich et~al.(2023)Stallrich, Allen-Moyer, and
  and}]{Stallrich02102023}
Stallrich, J., Allen-Moyer, K., and and, B.~J. (2023), \enquote{D- and
  A-Optimal Screening Designs,} \textit{Technometrics}, 65, 492--501.

\bibitem[{Stallrich and McKibben(2024)}]{Stallrich07082024}
Stallrich, J. and McKibben, M. (2024), \enquote{Optimal designs for two-stage
  inference,} \textit{Journal of Quality Technology}, 56, 327--341.

\bibitem[{Tang and Deng(1999)}]{G2}
Tang, B. and Deng, L.-Y. (1999), \enquote{Minimum {G$_2$}-aberration for
  nonregular fractional factorial designs,} \textit{The Annals of Statistics},
  27, 1914--1926.

\bibitem[{Wu and Hamada(2009)}]{WHtextbook}
Wu, C.~J. and Hamada, M.~S. (2009), \textit{Experiments: Planning, Analysis,
  and Optimization}, John Wiley \& Sons.

\bibitem[{Xiao et~al.(2012)Xiao, Lin, and Bai}]{xiao2012}
Xiao, L., Lin, D. K.~J., and Bai, F. (2012), \enquote{Constructing {D}efinitive
  {S}creening {D}esigns {U}sing {C}onference {M}atrices,} \textit{Journal of
  Quality Technology}, 44, 2--8.

\end{thebibliography}

\newpage
\section*{Supplementary Materials}

\subsection*{Proof $c(g_X) \to 1$ as $g_X \to \infty$}

\cite{Stallrich07082024} derived the expression
\[
c(g_X)=\sqrt{\frac{2}{g_X}} \ \frac{\Gamma\left(\frac{g_X}{2}+\frac{1}{2}\right)}{\Gamma\left(\frac{g_X}{2}\right)} \equiv \ \frac{\Gamma\left(g_X^*+\frac{1}{2}\right)}{\Gamma\left(g_X^*\right)\sqrt{g_X^*}}\ ,\
\]
where $g_X^* = g_X/2$. A known property of the $\Gamma$ function is that for $\alpha > 0$, 
\[
\lim_{z \to \infty} \frac{\Gamma(z+\alpha)}{\Gamma(z)z^\alpha} = 1\ ,\
\]
which implies $\lim_{g_X \to \infty} c(g_X)=1$ by setting $\alpha=1/2$.

\subsection*{Proof of Theorem~1}

Suppose $n=2(m+v)$ where $v \geq 1$ and $r(\Hmat)=m$. Then there are $v$ orthonormal vectors, $\V$, where $\Hmat^T\V=0$. For the foldover design, it follows that the $v$ vectors $(\V^T | -\V^T)^T$ will be orthonormal to $\X$ and so there will be $g_X \geq v$ degrees of freedom for unbiased estimation of $\sigma^2$. Next we show that there will be more than $v$ degrees of freedom if $\D$ contains replicates or foldovers. 

Group the rows of $\Hmat$ where the $g$-th group, denoted by $\Hmat_g$, has $r_g \geq 1$ replicated rows of some $\hvec_g$ and $f_g \geq 0$ folded over rows of $\hvec_g$. Without loss of generality, we can represent such a $\Hmat_g$ as:
\[
\Hmat_g=\begin{pmatrix}
    \hvec_g^T\\
    \vdots\\
    \hvec_g^T\\
    \hline
    -\hvec_g^T\\
    \vdots\\
    -\hvec_g^T
\end{pmatrix}=\begin{pmatrix} \onevec_{r_g}  \\ \hline -\onevec_{f_g} \end{pmatrix} \otimes \hvec_g^T
\]
Let $n_g=r_g+f_g$ be the number of runs in group $g$. Summing the $n_g$ values over groups yield $\sum_g n_g = m+v$. We reserve group $g=0$ to refer to all center runs, i.e., $\hvec_g=\zerovec$, and fix $f_0 = 0$. Let $G$ denote the number of groups having $\hvec_g \neq \zerovec$. Since $r(\Hmat)=m$, and an $\hvec_g=\zerovec$ does not affect the rank of $\Hmat$, it follows that $m \leq G \leq m+v$.  If $G=m+v$, then $n_0=0$ and $n_g=1$ for $g=1,\dots,G$. Otherwise we have at least one center run and/or some $g \geq 1$ with $r_g \geq 2$ or $f_g \geq 1$.

The foldover of $\Hmat$, $\D^\pm$, will also have the same group structure as $\Hmat$, but group $g$ will now have $2n_g$ runs. One may permute the rows of $\D^\pm$ to get the following representation
\[
\D^\pm=\begin{pmatrix}
    \phantom{-}\onevec_{2n_0} \otimes 0_m\\ \hline
    \phantom{-}\onevec_{n_1} \otimes \hvec_1^T \\ -\onevec_{n_1} \otimes \hvec_1^T \\ \hline 
    \vdots \\ \hline \phantom{-}\onevec_{n_G} \otimes \hvec_G^T \\ -\onevec_{n_G} \otimes \hvec_G^T
\end{pmatrix}
\]
This $\D^\pm$ would be produced for any $r_g\geq 1$ and $f_g \geq 0$ where $r_g+f_g=n_g$. Therefore, we may assume without loss of generality that $r_g=n_g$ and $f_g=0$, making $\Hmat_g=\onevec_{n_g}\otimes \hvec_g^T$. 

We now enumerate the $v$ null eigenvectors of $\Hmat$. If $n_0 > 0$, then $\Hmat_0=\zerovec$ exists and there is a matrix $\V_0$ comprised of $I_{n_0}$ in the first $n_0$ rows and 0's everywhere else where $\Hmat^T\V_0=\Hmat_0^T\I_{n_0}=\zerovec$. For a $g>0$ with $n_g>1$, let $\C_{n_g}$ denote a $n_g \times n_g-1$ matrix with columns comprised of orthonormal contrast vectors. Let $\V_g$ be the matrix which equals $\C_{n_g}$ in the rows corresponding to $\Hmat_g$ and $0$'s everywhere else, which satisfies $\Hmat^T\V_g=\Hmat_g^T\C_{n_g}=\zerovec$, as each column of $\Hmat_g$ is proportional to $\onevec_{n_g}$. Defining $R=\sum_{g (\geq 1)} (n_g-1)$, there must be $f=v-R-n_0$ remaining vectors in $\V$ that are orthogonal to all $\V_g$, $g \geq 0$. To be orthogonal to the $\V_1,\dots,\V_G$, the $f$ remaining vectors must have the property that their elements for rows corresponding to $\Hmat_g$'s with $n_g > 1$ must be a multiple of the $\onevec_{n_g}$ vector. To be orthogonal to $\V_0$, their corresponding elements in the $\Hmat_0$ runs must be 0. Denote the matrix of these remaining vectors by $\V_f$, where the $f$ is symbolic and should not be treated as a number, lest it be confused with one of the $\V_0,\dots,\V_G$.

Clearly the foldovers of $\V_0$, the existing $\V_g$ (i.e., where $n_g >1$), and $\V_f$ will be orthogonal to $\D^\pm$ and the corresponding $\X$. After applying the same row permutation as we did for $\D^\pm$ earlier, we get the following representations for the foldovers of $\V_0$:
\begin{align*}\V_0^\pm&=
\begin{pmatrix}
    I_{n_0} \\ -I_{n_0} \\ \hline 0 \\ 0 \\ \hline \vdots \\ \hline 0 \\ 0
\end{pmatrix}=\begin{pmatrix}
    I_{n_0}-n_0^{-1}J_{n_0} \\ 0 \\ \hline 0 \\ 0 \\ \hline \vdots \\ \hline 0 \\ 0
\end{pmatrix} + \begin{pmatrix}
    n_0^{-1}J_{n_0} \\ 0 \\ \hline 0 \\ 0 \\ \hline \vdots \\ \hline 0 \\ 0
\end{pmatrix} - \begin{pmatrix}
    0 \\ I_{n_0}-n_0^{-1}J_{n_0} \\ \hline 0 \\ 0 \\ \hline \vdots \\ \hline 0 \\ 0
\end{pmatrix}- \begin{pmatrix}
    0 \\ n_0^{-1}J_{n_0} \\ \hline 0 \\ 0 \\ \hline \vdots \\ \hline 0 \\ 0
\end{pmatrix}\ ,\\
&=\begin{pmatrix}
    I_{n_0}-n_0^{-1}J_{n_0} \\ 0 \\ \hline 0 \\ 0 \\ \hline \vdots \\ \hline 0 \\ 0
\end{pmatrix} -\begin{pmatrix}
    0 \\ I_{n_0}-n_0^{-1}J_{n_0} \\ \hline 0 \\ 0 \\ \hline \vdots \\ \hline 0 \\ 0
\end{pmatrix} + 
\begin{pmatrix}
    n_0^{-1}J_{n_0} \\ -n_0^{-1}J_{n_0} \\ \hline 0 \\ 0 \\ \hline \vdots \\ \hline 0 \\ 0
\end{pmatrix}\\
&\equiv \V_{0,1}^\pm - \V_{0,2}^\pm + \V_{0,3}^\pm\ .\
\end{align*}
For $g>0$ where $n_g > 1$, the foldover of $\V_g$ has the representation
\begin{align*}
\V_g^\pm&=\begin{pmatrix}
    0\\0 \\ \hline \vdots \\ \hline \phantom{-}\C_{n_g} \\ -\C_{n_g} \\ \hline 0 \\ 0 \\ \hline \vdots \\ \hline 0 \\ 0
\end{pmatrix}=\begin{pmatrix}
    0\\0 \\ \hline \vdots \\ \hline \C_{n_g} \\ 0 \\ \hline 0 \\ 0 \\ \hline \vdots \\ \hline 0 \\ 0
\end{pmatrix} - \begin{pmatrix}
    0\\0 \\ \hline \vdots \\ \hline 0 \\ \C_{n_g} \\ \hline 0 \\ 0 \\ \hline \vdots \\ \hline 0 \\ 0
\end{pmatrix} \equiv \V^\pm_{g,1}-\V^\pm_{g,2}\ .\
\end{align*}
It is straightforward to check that $\V_{0,1}^\pm$, $\V_{0,2}^\pm$, $\V_{0,3}^\pm$, $\V_f$, and all existing $\V_{g,1}^\pm$ and $\V_{g,2}^\pm$ are mutually orthogonal to each other, and are orthogonal to $\X$. Therefore, each matrix produces degrees of freedom for estimating $\sigma^2$, where the degrees of freedom equal the rank of the matrix. It follows that $\text{rank}(\V_{0,1}^\pm)=\text{rank}(\V_{0,2}^\pm)=n_0-1$ and $\text{rank}(\V_{0,3}^\pm)=1$. When $n_0 > 0$, the foldover of the center runs give $2(n_0-1)+1=2n_0-1$ pure error degrees of freedom. Next, $\text{rank}(\V_{g,1}^\pm)=\text{rank}(\V_{g,2}^\pm)=n_g-1$ so each $g$ with $n_g > 1$ produces $2(n_g-1)$ pure error degrees of freedom. Therefore, the pure error degrees of freedom for $\D^\pm$ will be
\[
p=\max(0,2n_0-1)+2\sum_{g\geq1}(n_g-1)\ ,\
\]
and the fake factor degrees of freedom will be $f=\text{rank}(\V_f)=v-R-n_0$.

\subsection*{Justification for $g_X=3$ or $4$}

\begin{figure}[h]
    \centering
    \includegraphics[width=0.5\linewidth]{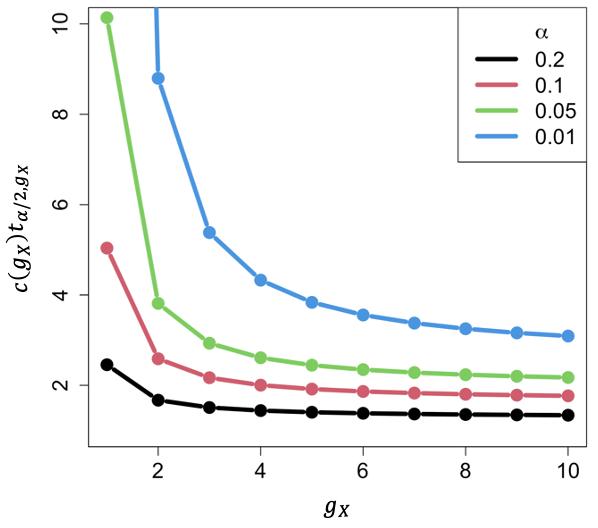}
    \caption{Plot of $c(g_X) \ t_{\alpha/2,g_X}$ as a function of $g_X$ for different $\alpha$ values.}
    \label{fig:SMFig1}
\end{figure}

We claim in sections~3.1 and 4 that $g_X=3$ or $4$ is sufficient for main effect inferences for small run designs. To explain, first note that $\lim_{g_X \to \infty} c(g_X) \ t_{\alpha/2,g_X} = z_{\alpha/2}$. Figure~\ref{fig:SMFig1} plots $c(g_X) \ t_{\alpha/2,g_X}$ as a function of $g_X$ for different $\alpha$ values. For $\alpha=0.10$ and $0.05$, the curves show the lower bounds are nearly attained when $g_X=3$ and $4$. Therefore, for large $n$ values we should avoid using complete foldover designs as they would lead to unnecessarily large $g_X$ values, as proven in Theorem~1.

\subsection*{Designs for $n=24$, $m=7$ Three Level Factors Example}

All seven designs considers in section~5.2 include a foldover design, with some also having augmented runs. These designs are displayed across Tables~\ref{tab:Section5_2_Designs_n24} to \ref{tab:Section5_2_Designs_n22}, based on the number of runs of the foldover design.

\begin{table}[h]
    \centering
    \caption{Half designs for complete foldover designs used in section~5.2. There are no augmented runs so each half design has $24/2=12$ runs.}
    \begin{tabular}{r|c|c|c}
         Name      & \textbf{ADSD.n24} & \textbf{R0.a05.n24} & \textbf{R1.n01.a05.n24}\\ \hline
         $\Hmat$   & $\begin{pmatrix} 
                     0 & + & + & + & + & + & + \\ \hline
                    + & 0 & + & - & + & + & + \\ \hline
                    + & - & 0 & + & - & + & + \\ \hline
                    + & + & - & 0 & + & - & + \\ \hline
                    + & - & + & - & 0 & + & - \\ \hline
                     + & - & - & + & - & 0 & + \\ \hline
                    + & - & - & - & + & - & 0 \\ \hline
                    + & + & - & - & - & + & - \\ \hline
                    + & + & + & - & - & - & + \\ \hline
                     + & + & + & + & - & - & - \\ \hline
                    + & - & + & + & + & - & - \\ \hline
                    + & + & - & + & + & + & -
                     \end{pmatrix}$ & 
                     $\begin{pmatrix} 
                     0  &  + &   + &  -  &  + &   +  & - \\ \hline
                     +  &  0 &  - &  -  & - &  -  & - \\ \hline
                    -  &  + &   0 &  -  & - &  -  & - \\ \hline
                     +  &  + &  - &   0  &  + &   +  & - \\ \hline
                    -  &  + &   + &   +  &  0 &  -  & - \\ \hline
                     +  &  + &  - &   +  & - &   0  &  + \\ \hline
                     +  & - &   + &  -  & - &   +  &  0 \\ \hline
                    -  & - &  - &   +  & - &   +  & - \\ 
                    -  & - &  - &   +  & - &   +  & - \\ \hline
                    -  & - &  - &  -  &  + &  -  &  + \\ \hline
                     +  & - &   + &   +  &  + &  -  & - \\ 
                    +  & - &   + &   +  &  + &  -  & -
                     \end{pmatrix}$ &
                     $\begin{pmatrix} 
                     0  &  0  &  0  &  0  &  0  &  0  &  0 \\ \hline
                     0  &  -  &  +  &  +  &  +  & -  &  + \\
                     0  &  -  &  +  &  +  &  +  & -  &  + \\ \hline
                     +  &  0  & -  &  +  & -  & -  & - \\
                     +  &  0  & -  &  +  & -  & -  & - \\ \hline
                     +  & -  &  0  & -  & -  & -  &  + \\ \hline
                     +  &  +  &  +  &  0  & -  &  +  &  + \\ \hline
                     +  & -  &  +  &  +  &  0  &  +  & - \\ \hline
                     +  &  +  & -  &  +  &  +  &  0  &  + \\ \hline
                     +  & -  & -  & -  &  +  &  +  &  0 \\ \hline
                     +  &  +  &  +  & -  &  +  & -  & - \\
                     +  &  +  &  +  & -  &  +  & -  & - \\
                     \end{pmatrix}$\\ 
    \end{tabular}
\label{tab:Section5_2_Designs_n24}
\end{table}

\begin{table}[h]
    \centering
    \caption{Half designs and augmented runs for designs in section~5.2 with $n=20$ runs in the foldover design. Each half design therefore has $20/2=10$ runs.}
    \begin{tabular}{r|c|c|c}
         Name      & \textbf{R0.a75.n20} & \textbf{R0.a05.n20} & \textbf{R1.n01.a05.n20}\\ \hline
         $\Hmat$   & $\begin{pmatrix} 
                     0  & -  & -  & -  &  +  &  +  &  + \\ \hline
                     +  &  0  &  +  &  +  &  +  &  +  &  + \\ \hline
                    -  & -  &  0  &  +  &  +  &  +  & - \\ \hline
                     +  &  +  & -  &  0  & -  &  +  & - \\ \hline
                    -  &  +  & -  & -  &  0  &  +  &  + \\ \hline
                    -  &  +  & -  &  +  &  +  &  0  & - \\ \hline
                    -  &  +  &  +  & -  &  +  & -  &  0 \\ \hline
                     +  &  +  & -  &  +  &  +  & -  &  + \\ \hline
                    -  &  +  &  +  &  +  & -  &  +  &  + \\ \hline
                    -  & -  & -  &  +  & -  & -  &  + \\
                     \end{pmatrix}$ & 
                     $\begin{pmatrix} 
                     0  &  +  & -  &  +  & -  &  +  &  + \\ \hline
                    -  &  0  & -  &  +  &  0  & -  & -    \\
                    -  &  0  & -  &  +  &  0  & -  & -    \\ \hline
                    -  &  +  &  0  & -  &  +  & -  &  + \\  \hline
                     +  &  +  & -  &  0  &  +  &  +  & - \\  \hline
                    -  &  +  &  +  &  +  &  +  &  0  &  + \\  \hline
                     +  &  +  &  +  &  +  & -  & -  &  0 \\
                     +  &  +  &  +  &  +  & -  & -  &  0 \\  \hline
                    -  & -  &  +  &  +  &  +  &  +  & - \\  \hline
                    -  &  +  &  +  & -  & -  &  +  & - \\
                     \end{pmatrix}$ &
                     $\begin{pmatrix} 
                     0  &  0  &  0 &   0  &  0  &  0  &  0 \\ \hline
                     0  & -  & + &  -  & +  & +  & + \\ \hline
                    +  &  0  & + &  +  & -  & +  & + \\ \hline
                    -  & -  &  0 &  +  & +  & -  & + \\ \hline
                    -  & +  & + &   0  & +  & +  & - \\ \hline
                    -  & +  & - &  +  &  0  & +  & + \\ \hline
                    +  & +  & - &  -  & +  &  0  & + \\ \hline
                    -  & -  & - &  -  & -  & +  &  0 \\ \hline
                    -  & +  & + &  -  & -  & -  & + \\
                    -  & +  & + &  -  & -  & -  & +
                     \end{pmatrix}$\\ \hline
         $\D_A$ & $\begin{pmatrix} 
         0 &   0 &   0 &   0 &   0  &  0 &   0 \\
         + &  - &  - &  - &  -  &  + &  - \\
         + &  - &  - &  - &  -  & - &   + \\
        - &  - &   + &   + &  -  & - &  -\end{pmatrix}$ & $\begin{pmatrix} 
         0  &  0  &  0  &  0  &  0  &  0  &  0 \\ 
        +  & -  & +  & +  & +  & +  & + \\ 
        -  & -  & -  & +  & +  & +  & + \\ 
        -  & -  & +  & -  & +  & +  & + \\ 
            \end{pmatrix}$ & 
         $\begin{pmatrix} 
         -  & +  & -  & +  & +  & -  & + \\
         -  & +  & +  & -  & -  & +  & + \\
         -  & +  & +  & +  & -  & -  & + \\
         -  & +  & -  & -  & +  & +  & +\end{pmatrix}$\\ \hline
    \end{tabular}
\label{tab:Section5_2_Designs_n20}
\end{table}


\begin{table}[h]
    \centering
    \caption{Half design and two augmented runs for design \textbf{SM.n22} in section~5.2 with $n=22$ runs in the foldover design. The half design has $11$ runs.}
    \begin{tabular}{r|c}
         Name      & \textbf{SM.n22} \\ \hline
         $\Hmat$   & $\begin{pmatrix} 
                        0&-&0&-& + &-&0\\ \hline
                        +  &  0 & 0 & - &  +  &  +  &  +  \\ \hline
                        - & 0 &  +  &  +  &  +  & 0 &  +  \\   \hline
                        
                      +  & - & 0 & - &  +  & 0 & - \\ \hline
                      +  & - & 0 &  +  & - &  +  &  +  \\ \hline
                      +  &  +  & 0 & - & - &  +  & -\\ \hline

                       +  &  +  & - & 0 & 0 & - &  +  \\ \hline
                      +  &  +  &  +  & 0 &  +  & - & - \\ \hline
                      -& + &-&0& + & + &-\\ \hline

                      + &-&-& + &0& + & + \\ \hline
                      
                      + &-&-& + & + &-& +

                     \end{pmatrix}$ \\ \hline
         $\D_A$ & $\begin{pmatrix} 
         0 & + & 0 & - & 0 & 0 & - \\
                                 0 & 0 & 0 & 0 & 0 & 0 & 0\end{pmatrix}$\\
    \end{tabular}
\label{tab:Section5_2_Designs_n22}
\end{table}

\end{document}